\documentclass[10pt,journal,compsoc]{IEEEtran}

\ifCLASSINFOpdf
  \usepackage[pdftex]{graphicx}

\else
 
\fi

\usepackage{amsmath}
\usepackage{algorithm}
\usepackage{algorithmicx}
\usepackage[noend]{algpseudocode}
\usepackage{enumitem}
\usepackage[utf8]{inputenc}
\usepackage[english]{babel}
 
\usepackage{amsthm}
 
\theoremstyle{definition}

\usepackage[font=footnotesize,caption=false]{subfig}

\setlist[itemize]{leftmargin=*}
\usepackage{xcolor}

\begin{document}

\title{Composing Energy Services in a Crowdsourced IoT Environment}

\author{Abdallah~Lakhdari,~\IEEEmembership{member,~IEEE,} Athman~Bouguettaya,~\IEEEmembership{Fellow,~IEEE,}
        Sajib~Mistry, and ~Azadeh~Ghari~Neiat
\IEEEcompsocitemizethanks{\IEEEcompsocthanksitem A. Lakhdari and A. Bouguettaya are with the School of Computer Science, The University of Sydney, Australia. E-mail: {abdallah.lakhdari, athman.bouguettaya}@sydney.edu.au.
\newline
S.~Mistry is with the School of Elect Eng, Computer and Math Sci, Curtin University, Australia. Email: sajib.mistry@curtin.edu.au.
\newline
A. Ghari~Neiat is with the School of Information Technology, Deakin University, Geelong, Australia. Email: azadeh.gharineiat@deakin.edu.au.\protect\\
}

}

\IEEEtitleabstractindextext{%

\vspace{-0.4 cm}
\begin{abstract}
We propose a novel framework for composing crowdsourced wireless energy services to satisfy users' energy requirements in a crowdsourced Internet of Things (IoT) environment. A new energy service model is designed to transform the harvested energy from IoT devices into crowdsourced services. We propose a new energy service composability model that considers the spatio-temporal aspects and the usage patterns of the IoT devices. A multiple local knapsack-based approach is developed to select an optimal set of partial energy services based on the deliverable energy capacity of IoT devices. We propose a heuristic-based composition approach using the temporal and energy capacity distributions of services. Experimental results demonstrate the effectiveness and efficiency of the proposed approach.
\end{abstract}
\vspace{-0.35 cm}
}

\maketitle

\IEEEdisplaynontitleabstractindextext

%
\IEEEpeerreviewmaketitle

\vspace{-1.1cm}
\IEEEraisesectionheading{\section{Introduction}\label{sec:introduction}}
\vspace{-0.2cm}
\IEEEPARstart{T}{he} concept of \textit{Internet of Things (IoT)} has emerged as a result of the advance in multiple technologies, including wireless communication, sensors, and embedded systems \cite{atzori2010internet}. Everyday \textit{things} are being transformed into IoT devices by embedding tiny sensors, actuators, computing resources and network connectivity. The smart IoT devices provide their augmented functionalities, e.g., sensing and actuating as \textit{IoT services}. This provides opportunities for integrating the \textit{physical} world with the \textit{cyber} world, enabling novel applications in several domains including smart cities, smart homes, agriculture, and healthcare.

\textit{Crowdsourcing} is an efficient way to leverage IoT services \cite{ren2015exploiting}. IoT users may crowdsource the functions of nearby IoT devices to fulfill their needs. Crowdsourcing enables a mobile ecosystem to share different services among IoT devices \cite{neiat2017crowdsourced}. For example, a set of co-located smartphones may provide computing or storage for a nearby resource-constrained smartwatch to render a map journey \cite{ahabak2015femto}. IoT devices can provide various types of crowdsourced services such as \textit{WiFi hotspots}, \textit{wireless charging} and \textit{environmental sensing} \cite{neiat2017crowdsourced,Previouswork11,perera2014sensing}. This provides opportunities to build novel applications such as spatio-temporal targeted recommender systems \cite{kefalas2018recommendations} and shared IoT service markets \cite{sun2016blockchain}.

The \textit{wireless energy transfer}, i.e, \textit{energy sharing as a service} is a key service in the \textit{dynamic} crowdsourced IoT ecosystem \cite{lu2016wireless}. It enables energy sharing between mobile IoT devices \textit{seamlessly}. 
The wireless energy sharing provides more \textit{convenience} to IoT users compared to carrying power banks or finding stationary power sources. Several IoT devices manufacturers have already adopted the wireless charging technology \cite{lu2016wireless}. For example, the inductive coupling for wireless energy transfer between two smartphones only allows a transmission within millimeters or centimeters \cite{cook2017wireless}. Significant research is striving to support a Watt-level energy transmission over a meter-level distance between IoT devices safely \cite{zhang2017adaptive}. Two companies, \textit{Energous}\footnote{https://www.energous.com/} and \textit{Wi-Charge}\footnote{https://wi-charge.com/} have already produced their charging devices prototypes which can deliver up to 3 Watts power over 5 meters to multiple receivers. Recently, the concept of {\em wireless crowd charging} has been introduced to provide IoT users with ubiquitous power access through crowdsourcing \cite{bulut2018crowdcharging,raptis2019online}.


We focus on \textit{crowdsourcing energy as a service} which has the potential to create a \textit{green} computing environment. The crowdsourced energy service ($CES$) has two main aspects, (a) \textit{harvesting energy}, and (b) \textit{the wireless transfer between IoT devices} \cite{bulut2018crowdcharging,dhungana2019peer}. The IoT devices are able to harvest energy from different natural sources, i.e., the kinetic movement of IoT users or their body heat \cite{gorlatova2014movers}. For example, a smart shoe may harvest energy from the physical activity of its user~\cite{choi2017wearable,lu2016wireless}. The harvested energy could be used to charge the nearby IoT devices \textit{wirelessly}. Energy providers might share energy altruistically to contribute to a green IoT environment. They can also be egoistic since energy is a vital resource for their IoT devices. Therefore, providers would not be interested in sharing their energy unless they receive a satisfying incentive to compensate for their resource consumption. There is a body of research that considers incentives for crowdsourced IoT services \cite{zhang2015incentives}. In this paper, we focus on composing crowdsourced energy services. We assume that the providers are incentivized by existing incentive models \cite{zhang2015incentives}.

The \textit{service paradigm} is a powerful mechanism to abstract the functionalities of IoT devices \cite{bouguettaya2017service}. We model the wireless energy transfer as a service. The \textit{function} of the wireless delivery of energy is represented formally along with its \textit{non-functional} attributes (i.e., Quality of Service (QoS)). The service abstraction enables some key operations such as the \textit{discovery of available energy services}, and the \textit{composition} of energy services based on users' requirements \cite{Previouswork11}. 

We focus on \textit{composing crowdsourced energy services} to fulfill an IoT users' energy requirements in \textit{confined areas} including coffee shops, restaurants, and waiting areas in airports. A single service provider may not be able to satisfy the energy requirements within the \textit{specified time interval}. In such a case, we need to combine multiple energy service providers. The composition of crowdsourced energy services is to select the optimal crowdsourced energy services that satisfy the user's energy and QoS requirements. The typical QoS of energy services include \textit{availability}, \textit{provision consistency}, and \textit{cost}. Note that, \textit{we only focus on the composition based on functionalities, i.e., fulfillment of energy requirements}. We also assume a \textit{static environment}. i.e., the user and providers \textit{do not move} during the composition.

To the best of our knowledge, there is a \textit{limited} research work on the composition of energy services. It is \textit{challenging to apply} the generic service composition approaches \textit{directly} \cite{chen2016goal} to compose the energy services due to the following characteristics of the crowdsourced IoT environment:
\vspace{-.09cm}
\begin{itemize}
    \item \textbf{Real-time composition:}
In pervasive computing, selecting a set of services across multiple mobile devices presents new challenges that do not occur in traditional settings in service computing \cite{chen2016goal}. Traditional service composition relies on previously generated composite services to build and complete future compositions. However, in the open environment of crowdsourced IoT, services are independently advertised, deployed and maintained by different IoT devices.  Reusing previously composed services is not always possible. 
\item \textbf{Partial invocation of energy services:}
One unique feature of the crowdsourced energy services is that they can be invoked \textit{partially}. An IoT user may consume only a part of the advertised energy from nearby services. 
\item \textbf{Wireless energy compatibility:}
Crowdsourced energy services require a \textit{novel composability model} of energy services. The energy services should consider \textit{intensity} compatibility between the user's IoT device and the providing devices in a composition. Note that, an IoT device may not receive more than a predefined recharging intensity \cite{lu2016wireless}.
\item \textbf{Spatio-temporal service discovery:}
Special considerations are necessary for the efficiency and performance of the selected set of services \cite{neiat2017crowdsourced}. In particular, in the crowdsourced IoT environment.  IoT services providers and consumers have different spatio-temporal preferences. Therefore, the selected set of crowdsourced energy services will not fulfil the requirement of a consumer if these services are not composable according to: (i) their spatio-temporal features and (ii) preferences of the energy consumers. 
\item \textbf{Inconsistent provision of crowdsourced energy services:} 
Crowdsourced energy services deliver energy wirelessly. Wireless communication channels are sensitive to the distance and sometimes unreliable. Additionally, the crowdsourced energy services are provided from IoT devices which are already in use by their owners. Hence, delivering consistent wireless energy from an IoT device to another depends on the usage of the device owners.
\item \textbf{Volatile crowdsourced energy services:} 
Service providers establish a wireless network in ad hoc ways. energy service provision relies mainly on the distance between IoT devices. They may offer and drop services at arbitrary times due to their mobility. Predicting the availability of crowdsourced energy services depends on defining the context (i.e., location and time of the day) in addition to the usage behavior of the IoT user \cite{cardone2014participact}. 

\end{itemize}

We propose a composition framework of energy services extending the energy service model proposed in \cite{Previouswork11}. The proposed framework considers all the aforementioned challenges except the volatile behavior of energy services due to the mobility of the crowd.  In this paper, we focus on composing static services. In the future, we will consider composing mobile energy services in a crowdsourced IoT environment. The aim of this paper is to answer the questions: (a) \textit{how to model energy services and queries in a crowdsourced IoT environment?}, (b) \textit{what are the composability rules for crowdsourced energy services?}, and (c) \textit{ how to compose nearby crowdsourced energy services to fulfill the energy requirement of an IoT user?}. The \textit{service model} enables the device owners to \textit{advertise} the harvested energy as services. The \textit{query model} allows the spatio-temporal service discovery and composition for nearby consumers. The \textit{composability model} defines the possible candidate services for the composition. Here,  we assume that \textit{the no-lock in contract service invocation is allowed}, i.e., the user can \textit{leave a service any time} during the advertised service time. We propose a \textit{modified} temporal knapsack algorithm \cite{bartlett2005temporal} to compose crowdsourced energy services. The composition technique considers the energy \textit{description} attributes and the \textit{spatio-temporal} features of IoT devices to select and compose energy services. The main contributions of this paper are: 
\begin{itemize}
    \item Designing a novel composability model for crowdsourced energy services.
    \item Developing an \textit{energy-usage aware} Quality of Service (QoS) model to evaluate energy services.
    \item Developing a \textit{heuristic-based} spatio-temporal composition approach to select the best composition of energy services. 
    \item Conducting experiments on real datasets to illustrate the performance and effectiveness of the temporal composition approach.
\end{itemize}



\IEEEpeerreviewmaketitle
\vspace{-0.5 cm}
\section*{Motivation scenario}
\vspace{-0.1 cm}

		\begin{figure}[!t]
		    \centering
		    \includegraphics[width=.50\textwidth]{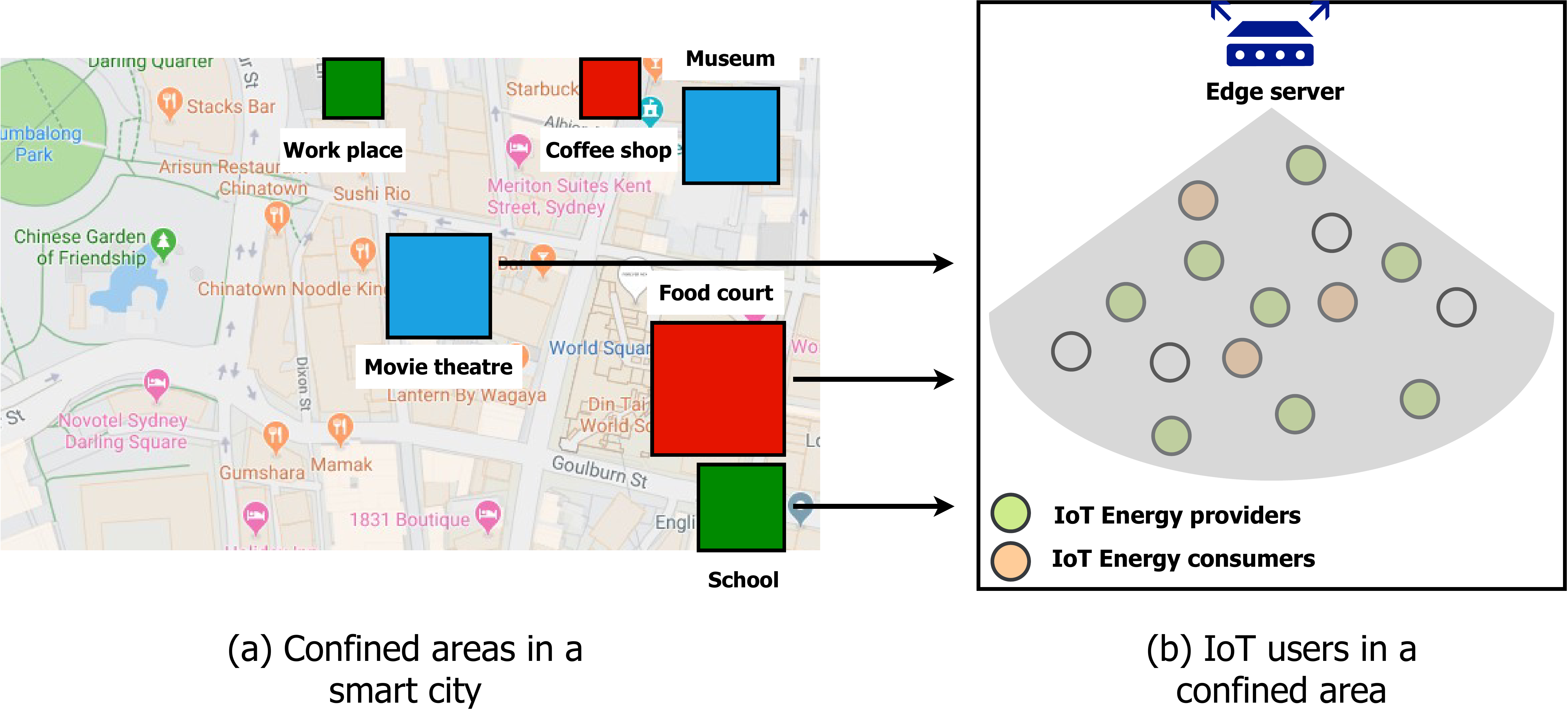}
        \caption{\small{Crowdsourced energy services in confined areas}}
        \label{fig:microcell}
        \vspace{-0.4cm}
		\end{figure}
\begin{figure} [!t]
		    \centering
		    \includegraphics[width=.35\textwidth]{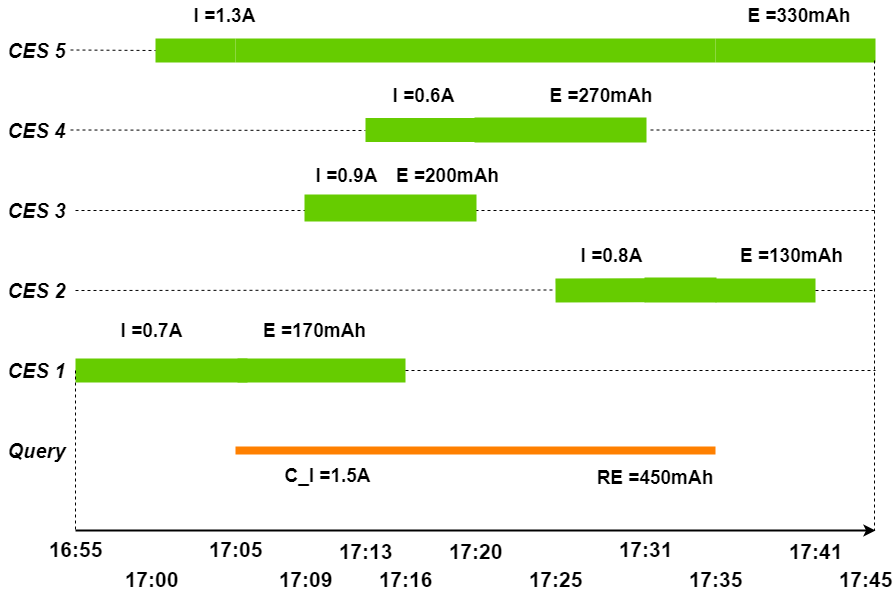}
        \caption{\small{Services time intervals for an IoT user \textit{X}'s  query $Q$}}
        \label{fig:meta1}
        \vspace{-0.4cm}
		\end{figure}

People may gather in different places (i.e., confined areas) in a smart city e.g., coffee shop, restaurant, workspace, theatre, etc. (see figure \ref{fig:microcell}a). They may harvest energy by their wearables \cite{TranM0B19}. They may also share their spare energy wirelessly with nearby IoT devices. The distance between IoT devices exchanging energy may reach five meters to ensure a successful wireless transmission. The IoT devices and wearables are assumed to be equipped with wireless energy transmitters and receivers e.g., Energous  and Wi-Charge. Let us consider a typical scenario of an IoT user in a coffee shop. An IoT user `\textit{X}' wants to charge the depleting battery of its smartphone in a coffee shop. The user `\textit{X}' needs to run some critical applications e.g., to make a call and to check for upcoming appointments. The energy query is processed at the edge i.e., a router in a confined area (see Figure \ref{fig:microcell}b). All local energy queries and advertised energy services are processed at the edge associated to their confined area. The smartphone casts an energy query in the following way \textit{the user $X$ is looking for an amount of energy $E =450mAh$ in the location $l$ during the time $[t_1=17:05~,~t_2=17:35]$}. Five people in the same coffee shop are willing to share their energy during \textit{X}'s query period. The required energy can be provided from the harvested energy by wearables of these people or from their smartphone batteries. Each crowdsourced energy service \textit{CES} is defined by its available time and provided energy (see Figure  \ref{fig:meta1}). None of the available services can provide the required amount of energy to `\textit{X}' as all available energy services provide less than $450mAh$. As a result the composition of multiple services is required which may provide the required amount of energy. However, the composition must consider the compatibility of the  received current intensity, i.e., the intensity of the aggregated received current cannot exceed the predefined compatibility intensity of the consuming device, if multiple energy services are providing energy at the same time. Hence, the composition should select the \textit{optimal combination of services}. For example, selecting $CES~5$ for the entire query interval prevents the composition of other energy services like  $CES~2$, $CES~3$, and $CES~4$, because of the intensity compatibility condition. As the no-lock in contract service invocation is allowed and the composability of component services, chunking the query duration and  composing the partial invocations of $CES~1$, $CES~3$, and $CES~4$ given in Table \ref{tab:table5} can fulfill the required $450mAh$ energy requirement.

        \begin{table}[ht]
		\caption{ Energy services for an IoT user \textit{X}'s  query $Q$}
		\label{tab:table5}
		\centering
	\begin{tabular}{|l| l| l| l| c|} 
 			\hline
            			\footnotesize{ }& \footnotesize{\textbf{Chunk 1}} & \footnotesize{\textbf{Chunk 2}} & \footnotesize{\textbf{Chunk 3}}  & \footnotesize{\textbf{Energy}} \\
 			\hline
			$C_{1}  $ & $\{CES_{5} \}$ & $\{CES_{5} \}$ & $\{CES_{5} \}$& $ 330 $ \\
			\hline
			$C_{2} $ & $\{CES_{1} \}$ & $\{CES_{1}, CES_{3} \}$& $\{CES_{3}, CES_{4} \}$ & $450 $\\
 			\hline
		\end{tabular}
		 \end{table}  
		


\vspace{-0.6 cm}
\section{The Framework for Composing Crowdsourced Energy Services (CES)}\label{sysmdlpb}
\begin{figure}[!t]
\begin{center}
\includegraphics[width=0.4\textwidth,height=3.6cm]{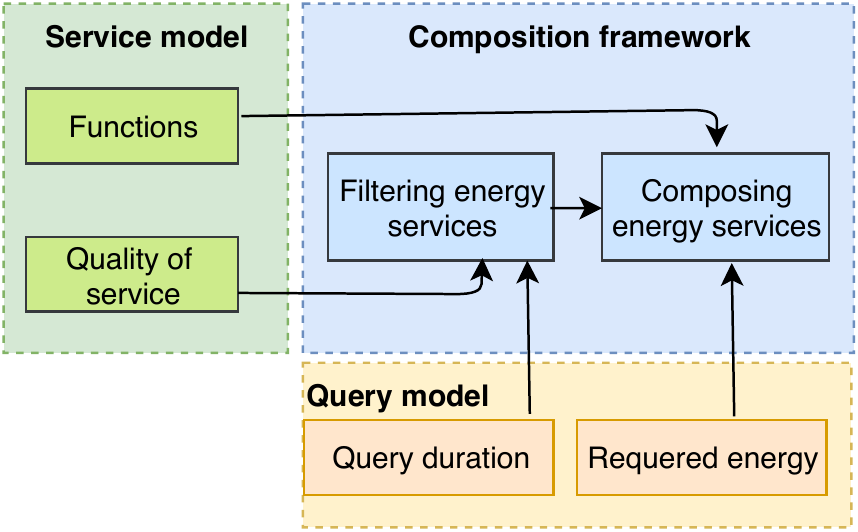}
\caption{\small{Framework for composing Crowdsourced Energy Services}}
\label{fig:FRMWRK}
\end{center}
\vspace{-0.35 cm}
\end{figure}
Let us assume, in a confined area $A$, there exists a set of $n$ energy services $S=\{S_1,~S_2,~ \dots,~ S_n\}$ and a set of $m$ energy queries $Q=\{Q_1,~Q_2,~ \dots,~ Q_m\}$. Each service is represented by the provided amount of energy $S_j.EN$ and the start time $S_j.st$ and end time  $S_j.et$. The energy queries also are described by the required amount of energy $Q_i.RE$ and the start and the end time of the energy query, $Q_i.st$ and $Q_i.et$ respectively. \textit{We formulate crowdsourcing energy services into a service composition problem}. Composing energy services in a crowdsourced IoT environment needs to consider the spatio-temporal features of services and queries. Providing energy for a query $Q_i \in Q$ requires the composition of services $S_j \in S$ where $[S_j.st~,~S_j.et] ~\subset~[Q_i.st~,~Q_i.et] $ and $\Sigma S_j.EN \geq Q_i.RE$. We consider the following \textit{assumptions} about a confined crowdsourced IoT environment:
\begin{itemize}
    \item The IoT devices in the crowdsourced environment are equipped  to receive and transmit energy wirelessly. 
    \item The energy service providers are fixed in one location inside the confined area for the whole duration of the energy query.
    \item The selected services deliver energy continuously once invoked until the consumer finishes. The invocation could be finished by switching to another service or the service is closed by the provider.
    \item The energy consumers may receive energy from multiple providers at the same time if their aggregated current intensity is lower or equal to the compatibility intensity of the consuming device.
    \item It is possible to invoke the energy services partially. The crowdsourced energy services do not have any Service Level Agreement (SLA) and the services do not have any lock in contract. 
    \item The providers have fluctuating energy usage behavior.
    
\end{itemize}

We identify four main components to build a composition framework for crowdsourced energy services (CES): 
\begin{itemize}
    \item \textbf{Crowdsourced energy service model:} The service model describes the function of delivering energy wirelessly and the associated qualities. This representation facilitates implementing a platform for crowdsourcing energy in an IoT environment. Providers use this model to advertise their energy services.

    \item \textbf{Energy query:} The energy query model is defined to describe the IoT users energy requirements and preferences in the simplest way. A query model represents the spatio-temporal preferences and the energy requirements of an energy consumer. The energy query $Q_i \in Q$ is the main input to the framework. It defines the filtering parameters for the proposed \textit{query dependent composability model}.  
    \item \textbf{Filtering crowdsourced energy services:} The composability model defines whether two energy services are composable according to an energy query. This model uses the query  spatio-temporal features to find the available nearby energy services. Crowdsourced energy services are filtered based on the spatio-temporal features and the consuming IoT device features. 
    \item  \textbf{Composing crowdsourced energy services:} The composition algorithm component finds the optimal composition of energy services which provides the required amount of energy within the query duration. The filtered energy services are composed to provide the required amount of energy to the consumer. The proposed composition algorithm is an extension of the temporal $0/1$ knapsack algorithm. the aim of this extension is   to improve the performance and the time consumption of the temporal $0/1$ knapsack algorithm.    
\end{itemize}
\vspace{-0.5 cm}
\section{Crowdsourced Energy Service model}
\vspace{-0.1 cm}
Crowdsourcing energy from IoT devices relies mainly on sharing IoT services with accessible mobile devices in the proximity. IoT-based energy services are modelled using the spatio-temporal aspects of device owners \cite{neiat2017crowdsourced}. We extend the existing energy service model \cite{Previouswork11} by \textit{considering the dynamic energy usage behavior of providers and consumers}. We formally define the crowdsourced energy service model as follows.       


\textbf{Definition 1}: A Crowdsourced Energy Service
$ CES $ is a tuple of $< Eid, Eownerid, F , Q >$ where:
\begin{itemize}
\item $Eid$ is a unique service ID,
\item $Eownerid$ is the unique ID of the IoT device owner,
\item $F$ is the set of  $CES$  IoT device's functionalities. 

\item $Q$ is a tuple of $<q_1, q_2, ..., q_n>$ where each $q_i$ denotes a QoS property of $CES$, e.g., energy capacity. 

\end{itemize}

The energy usage behavior of IoT devices also needs to be modeled to ensure a consistent provision of energy services. The energy capacity of IoT devices changes over time and depends on the user consumption behavior \cite{Carroll2010AnAO}. Several consumption models have been proposed for IoT devices \cite{oliver2011empirical} \cite{church2015understanding}. The energy consumption behavior is represented as a time series of the state of charge $SoC$ by a timestamp $\{~SoC(t): t \in T~\}$.

We also need to capture the regularity of the energy usage behavior of the device. We use $Kolmogorov-Sinai~~entropy$ \cite{latora1999kolmogorov} to define the regularity of energy usage time series. The lower the \textit{entropy} value, the more \textit{regular} behavior of energy usage by the IoT device is.  
We use the entropy to define some QoS attributes of an energy service.  


\begin{itemize}
	
	 \item \textbf{Energy Intensity}: Energy Intensity represents the intensity of the wirelessly transferred energy. The energy is transferred under a certain voltage. We assume that all IoT devices that are related to energy services are functioning under a voltage between 3 and 5 volts. These IoT devices are also compatible in term of voltage.   
     \item \textbf{Transmission success rate}: Transmission success rate $Tsr$ is the ratio between the transmitted energy from the energy provider and the received energy by the energy consumer ~\cite{he2013energy}. $Tsr$ is calculated as follows.
\begin{equation}
\footnotesize
Tsr = \frac{G_{t} G_{r} \gamma}{L_{p}} \left( \frac{\lambda}{4\pi(D+\beta)} \right)^{\theta}    
\end{equation}    
where $G_t$, $G_{r}$, $L_{p}$,$\gamma$, $\lambda$, $\beta$, $\theta$, and $D$ represent the transmission gain, the reception gain, polarization loss, rectifier efficiency, wave length, short distance energy transmission parameter, path loss coefficient, and the distance between devices, respectively. 

\item \textbf{Provision consistency parameter}: The consistency parameter $\alpha$ is calculated by the formula 
\begin{equation}
\small
  \alpha = \frac{1}{Kolent}~.~EUB~ 
\end{equation}


where $Kolent$ is the approximate entropy. If the energy usage behavior of the device is regular, $\alpha$ increases significantly. The usage time series shows a seasonal behavior which increases the  $\frac{1}{Kolent}$ value. $EUB $ is the energy usage behavior of the device owner. In \cite{Carroll2010AnAO},  different usage patterns of IoT devices are defined. According to these patterns, the energy consumption of the device can be estimated. Providers who want to share their energy have to follow one of these patterns as follows:  \textit{Suspend} i.e., not using their devices e.g., $EUB = 1$,  \textit{Casual} i.e., using them casually with
few functionalities e.g., $EUB = 0.75$, or \textit{Regular} i.e., using them with a predictable usage behavior e.g., $EUB = 0.50$. 
\item \textbf{Deliverable Energy capacity}: Deliverable Energy capacity, $DEC$ is the energy capacity that a consumer realistically receives. It is affected by the \textit{Transmission success rate} $Tsr$ and the \textit{provision consistency parameter} $\alpha$. $DEC$ is calculated from the advertised $EC$ as follows which is given by \textit{milliAmpere hour} $mAh$.   
\begin{equation}
\footnotesize    DEC= \alpha~.~EC~.~Tsr
\end{equation}
    

\item \textbf{Location}: Location, $loc$ is the GPS location of the IoT device providing an energy service.

\item \textbf{Start time}: Start time $st$ is the time of launching an energy service by an IoT device. It is assumed to be announced by energy service providers.

    
\item \textbf{End time}: Given the initial energy capacity $EC$, the intensity of the transferred current $I$, and the start-time of the energy service $st$, the service effective end time $ et$ can be estimated by the following formula:
   \begin{equation}
 \footnotesize  
  et= st + ~\frac{EC}{I}    
   \end{equation}
   \end{itemize}

\vspace{-0.3 cm}
\subsection*{The energy query model}
\textbf{Definition 2}: A Crowdsourced   Energy Service Consumer Query is defined as a tuple $Q< t, l, Re, I, d, Cl>$:  
\begin{itemize}
	\item $t$ refers to the timestamp when the query is launched.
	\item $l $ refers to the location of the energy service consumer. We assume that the consumer stays fixed after launching the query.
	\item $RE$ represents the required amount of energy. 
	\item $I$ is the maximum intensity of the wireless current that a consuming IoT device can receive. 
	\item $d$ refers to a user-defined charging period of time designated by its start time and end time. 
	\item $Cl$ is the coordination loss. It is the amount of energy to be spent for the connection establishment between the consumer and the provider.

\end{itemize}

\textbf{Definition 3}: Given a set of crowdsourced energy services $S_{CES}= \{ CES_{1}, CES_{2}, \dots CES_{n} \}$ and a query $Q< t, l, Re, I, d, Cl>$, the spatio-temporal crowdsourced energy service composition problem is formulated as selecting the optimal composition of nearby  energy services $ CES_{i} \in S_{CES}$ that can transfer the maximum amount of energy from the nearby IoT devices in the \textit{shortest} and the \textit{earliest} time period.



\vspace{-0.4 cm}
\section{Composability model of CES}
We present a composability model of crowdsourced energy services to check whether two services are composable according to the constraints of the crowdsourced IoT environment. Four new composability rules are defined based on the spatio-temporal constraints of IoT users, the interface heterogeneity of IoT devices, and the energy loss features. We define \textit{query dependent composability} rules according to the consumer's spatio-temporal preferences. \textit{The IoT device-related composability} rules are defined by \textit{intrinsic properties} of the IoT device's characteristics and interface. Examples of intrinsic properties include the intensity of the wireless energy and the energy loss while establishing a connection between two IoT devices. We define four composability rules as follows:

(a) \textbf{Spatial composability}:
Given two crowdsourced energy services $CES_i$ and $CES_j$ and query location $Q.l$, $CES_i$ and $CES_j$ are spatially composable, if and only if the distance between the location $l$ of each energy service and the location  $Q.l$ is less than the distance permitting a successful energy wireless transfer $ESD$. Equation \ref{spatial} describes the Spatial composability. \textit{We assume $ESD$ is fixed for all human-centric IoT devices}.   
\begin{equation}
\label{spatial}
\small   D(CES_i.l~,~ Q.l)\leq ESD ~~and~~ D(CES_j.l~,~ Q.l)\leq ESD 
\end{equation}

\begin{figure}[!t]
\centering
\includegraphics[width=0.35\textwidth,height=3.5cm]{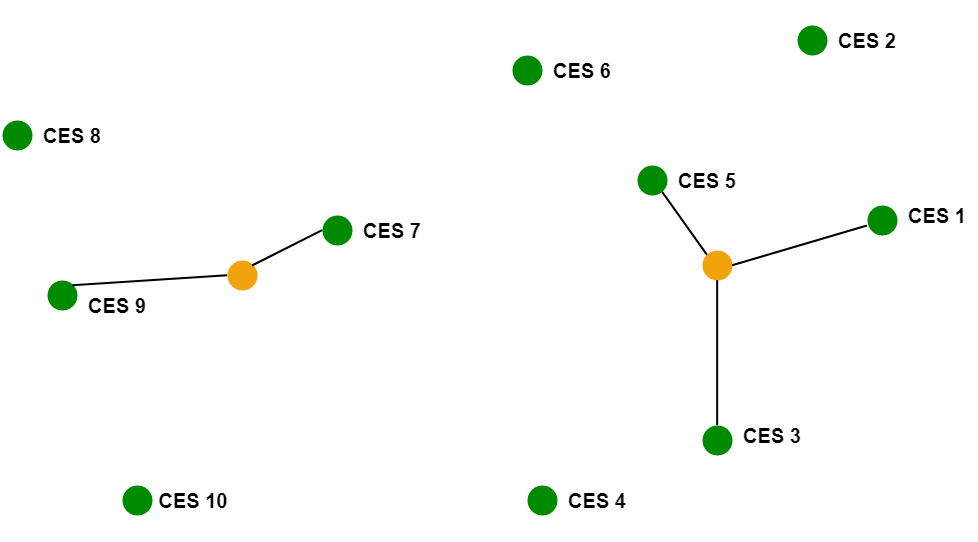}
         \caption{\small{Spatial composability of energy services}}
         \label{fig:Space_compo}
         \vspace{-0.25cm}
\end{figure}

In Equation \ref{spatial}, $D$ refers to the Euclidean distance. For example, in Figure \ref{fig:Space_compo}, $CES_1$, $CES_3$, and $CES_5$ are composable for the first query, on the right ,  and $CES_7$ and $CES_9$ are composable for the second query, on the left.

(b)\textbf{Temporal composability}:
The energy query duration defines whether two energy services are temporally composable. We use Allen's interval algebra \cite{allen1990maintaining} to define the temporal composability of energy services. \textit{Interval algebra} defines services and query duration as intervals delimited by their start and end time $<Start~time~,~End~time>$. If $CES_i$, $CES_j$, and $Q$  are two crowdsourced energy services and an energy query respectively, $CES_i$ and $CES_j$ are temporally composable, if and only if the duration of each service is  within the query duration $Q.d$ . 

We also define \textit{partial temporal composability} for energy services. If the duration of an energy service $CES_i$ only overlaps with the query duration $Q.d$ in the beginning or at the end, the energy service is considered as a partially available service. We derive a new service called partial service $CES_{i}^\prime$ by only the duration which is within $Q.d$. QoS parameters of $CES_{i}^\prime$ have to be recalculated according to the new duration start and end time. For example, if $CES_{i}$  overlaps with the query duration in the beginning, the \textit{start time} $CES_{i}^\prime.st$ will be the query launching time $Q.t$. The new value of the deliverable energy capacity of this service is recalculated using Equation \ref{temporal} because we consider a uniform distribution in terms of energy consumption for the device and delivery. 
\begin{equation}
\label{temporal}
\small
\footnotesize  CES_i^\prime.DEC = (CES_i.et-Q.t)~.~CES_i.I~.~CES_i.Tsr~.~CES_i.\alpha  
\end{equation}


In Equation \ref{temporal}, $CES_i.st$ and $CES_i.et$ are the new start time and end time of $CES_i$ respectively. $CES_i.I$ is the intensity of the wirelessly transferred energy. $CES_i.Tsr$ is the wireless transmission success rate and $CES_i.\alpha$ is the \textit{provision consistency} parameter of the service.

\begin{figure}[!t]
    \centering
    \includegraphics[width=0.35\textwidth]{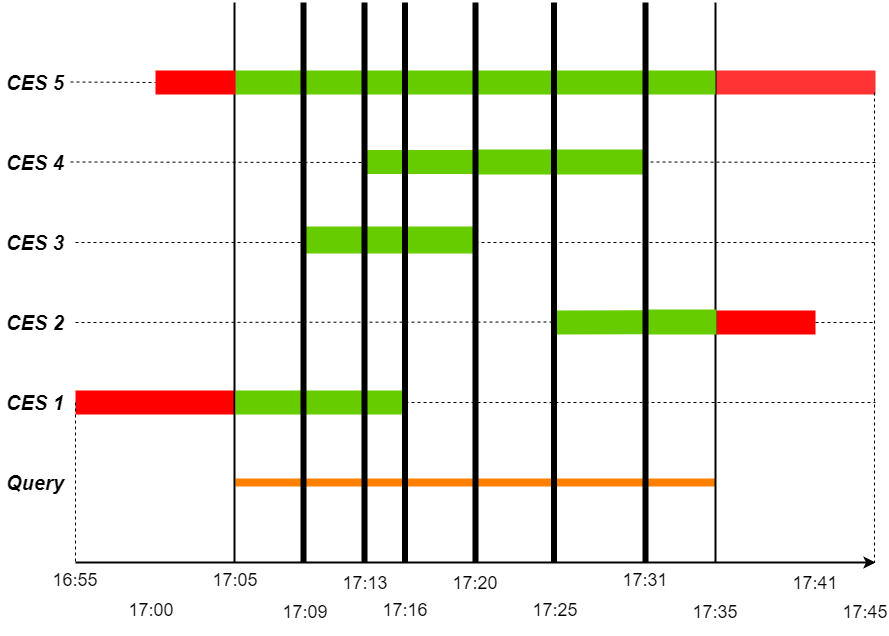}
        \caption{\small{Temporal representation of energy services}}
        \label{fig:Examplechunking}
        \vspace{-0.4cm}
\end{figure}

(c) \textbf{Intensity compatibility}:
There are two different rules to check the current compatible composability of energy services according to the composition scenario: \textit{simultaneous} or \textit{sequential}. In both scenarios, the received energy should not be higher than the maximum intensity supported by the consuming IoT device. Given two crowdsourced energy services $CES_i$ and $CES_j$ and an energy query $Q$, the intensity compatibility is defined as follows:
\begin{itemize}
    \item $CES_1$ and $CES_2$ are sequentially composable if: 
    \begin{equation*}
        \footnotesize{CES_1.I\leq Q.I~~and~~CES_2.I\leq Q.I}
    \end{equation*}
    \item $CES_1$ and $CES_2$ are simultaneously composable if:
    \begin{equation*}
        \footnotesize{CES_1.I+CES_2.I\leq Q.I}
    \end{equation*}
\end{itemize}

(d) \textbf{Composition Eligibility}:
The energy services from human-centric IoT devices may scale from very \textit{small amounts} (provided by tiny IoT devices) to a \textit{considerable amounts} shared by bigger devices. In this regard, a consumer might spend more energy to receive than the provided small amount of energy from tiny energy services. \textit{The energy could be spent in the service discovery and for the connection establishment between the consumer and the provider}~\cite{na2018energy}. Hence, some services may not be fit according to the user queries. it may not be possible to switch from a service to another only after a minimum connection time which allows provisioning energy amount higher than what has been spent in the connection establishment. We define this phenomena as \textit{coordination loss}. We calculate the coordination loss of a query $Q.Cl$ in Equation \ref{coloss}
\begin{equation}
\label{coloss}
\footnotesize
    Q.Cl = C_c~.~f_c + C_i~.~f_i
\end{equation}
In Equation \ref{coloss}, $C_c$ is the rate of energy consumption for communicating one unit of data with the cloud. $f_c$ is the flow of data interchanged between the consuming device and the cloud. Similarly, $C_i$ is the rate of energy consumption for communicating one unit of data with the IoT device providing the energy service $i$. $f_i$ is the flow of data interchanged between the consuming device and the IoT device providing the service $i$.  
The service provider also spends energy to coordinate the wireless energy transfer with the consuming device. We call this energy amount $PCl$ \textit{Provider's Coordination Loss}. $PCl $ is defined in the same way with $Q.Cl$ in Equation \ref{pcl}.
\begin{equation}
\footnotesize
\label{pcl}
  PCl = C_c~.~f_c + C_j~.~f_j  
\end{equation}

In Equation \ref{pcl}, $C_j$ and $f_j$ are the rate of energy consumption for communicating one unit of data with the IoT device requesting energy  $j$ and the flow of data interchanged between the consuming device $j$ respectively. An energy service $CES$ is a component service for an energy query $Q$ if and only if the coordination loss for the query $Q.Cl$ is lower than the provided energy by the service $CES$:
\begin{equation}
\footnotesize
 Q.Cl << (CES.DEC~.~CES.Tsr-CES.PCl)   
\end{equation}

\vspace{-0.35 cm}
\section{Composition of CES}
In this section, we present the composition framework of crowdsourced energy services. The aim of the composition is to { \em maximize the obtained energy by the consumer within the query duration from the available nearby services  with respect to their spatio-temporal and energy related constraints}. We start by explaining the filtering process of energy service candidates based on their spatio-temporal features. We present then different spatio-temporal composition techniques of crowdsourced energy services. Finally, we propose a heuristic to reduce the search space of candidate energy services. 

\vspace{-0.3 cm}
\subsection{Filtering crowdsourced energy services}
The \textit{first step} in the composition is to design the filtering process to select the candidate services. 
We need an efficient \textit{indexing method} for the fast discovery of energy services. The \textit{location} and \textit{time} are intrinsic parts of energy services. Therefore, we index energy services based on spatio-temporal characteristics. The \textit{3D R-tree} is a spatio-temporal index data structure which deals with range queries of the type ``\textit{report all objects within a specific area during the given time interval}''~\cite{jun2003dynamic}. The time is added as the third axis to spatial axes. When a query $Q$ arrives, an area is defined by the location of the consumer $Q.l$ and a distance  allowing the wireless power transmission between IoT devices $r$. We use \textit{3D R-tree} to index services spatio-temporally.

We deploy a function called \textit{Spatio-temporal selection algorithm} (see Algorithm~\ref{euclid}) to select energy services spatio-temporally \cite{neiat2017crowdsourced}. The function takes input parameters as the position of the consumer $Q.l$, the start time of the query $Q.t$, and the duration of the query $Q.d$. The output of \textit{Spatio-temporal selection algorithm} is the set of all nearby available services $NearbyS$ between the start time and the end time of the query (Algorithm~\ref{euclid}-Lines 1,2). We select services located in a defined area at the time interval $[Q.t, Q.t+d]$ using the spatio-temporal composability rules. Each energy service has a time interval $[st,et]$ and a location $loc$. Figure ~\ref{fig:Examplechunking} represents a query $Q$ and five energy services $ CES $. 

The services are filtered spatially by selecting just the services inside the area defined by the consumer location $Q.l$ and the \textit{spatial} composability rule (see Figure \ref{fig:Space_compo}). The services are also filtered temporally by the \textit{temporal} composability rule, choosing just services having time interval within or overlapping with the query duration $[Q.t, Q.t+d]$ (see Figure \ref{fig:Examplechunking}). Each leaf in the \textit{3D R-tree} is considered as $CES$. A search cube $SC$ is determined by the query location $Q.l$ and duration $Q.d$. All leaves inside or which overlap with the search cube are selected (Algorithm~\ref{euclid}-Lines 3-7). The overlapping services with the query duration like $CES1$, $CES2$, and $CES5$ in Figure \ref{fig:Examplechunking} are called \textit{partially available services}. The query duration overlaps only with parts of the time intervals of these services. We define new services only in the parts overlapping with the query duration. We also consider the provided energy services only in the overlapping parts (Algorithm~\ref{euclid}-Lines 9-23).  
\begin{algorithm}[!t]
\footnotesize
	\caption{ Spatio-temporal selection algorithm}\label{euclid}
	\begin{algorithmic}[1]
		\State \textbf{Input: }$ Q.l$, $Q.t$, $Q.d$
		\State \textbf{Output: }$NearbyS$ \text{~// Nearby services during $Q.d$}
		\State $SC=$ \text{Compute search cube based on $Q.l$, $Q.t$, and $Q.d$}
		\State \text{// Lower bound of $SC=[(Q.l.x-r)*cos(45)],$ }
		\State \text{//$~~~~~~~~~~~~~~~~[(Q.l.y-r)*sin(45)], Q.t$ }
		\State \text{// Upper bound of $SC=[(Q.l.x+r)*cos(45)],$ }
		\State \text{// ~~~~~~~~~~~~~~~~ $[(Q.l.y+r)*sin(45)], Q.t+d$ }
		\State $NearbyS = \emptyset$
		\State \textbf{For all $ CES_i \in SC$ }
		\If {($CES_i.st\geq Q.t$ \text{and } $CES_i.et\leq Q.t+d $)}  
		\State \text{$NearbyS = NearbyS \cup \{CES_i\}  $}
		\State  \textbf{else}
		\If {($CES_i.st< Q.t$ \text{ and } $CES_i.et\leq Q.t+d $) \text{ or } ($CES_i.st \geq  Q.t$ \text{ And } $CES_i.et> Q.t+d $)}  
		\State  $CES_i.st\gets Q.t $   
        \State  \text{Or }
        \State $CES_i.et\gets Q.t+d $   
		\State  $CES_i.DEC\gets (CES_i.et-CES_i.st)*CES_i.I*CES_i.Tsr $ 
		\State \text{$NearbyS = NearbyS \cup \{CES_i\}  $}
		\State  \textbf{else}
		
		\State  $CES_i.st\gets Q.t $ 
		\State  $CES_i.et\gets Q.t+d $  
		\State  $CES_i.DEC\gets (CES_i.et-CES_i.st)*CES_i.I*CES_i.Tsr $
		\State \text{$NearbyS = NearbyS \cup \{CES_i\}  $}	
		\EndIf 
		\EndIf
		
		\Return $ NearbyS $
	\end{algorithmic}
\end{algorithm}



Let us assume that given two services $CES_i$ and $CES_j$ and an energy query $Q$, the availability time of $CES_i$ and $CES_j$ are within a query duration $Q.d$. The waiting time between the two services $CES_i$ and $CES_j$, called $Wt$, is defined in Equation \ref{waittime}:
\begin{equation}
\footnotesize
\label{waittime}    
Wt_{ij} = CES_j.st - CES_i.et
\end{equation}

In Equation \ref{waittime}, $CES_i.et$  and $CES_j.st$  are end time and start time of $CES_i$ and $CES_j$ respectively. 
$CES_i$ should not be considered as a component service for the energy query $Q$ if and only if the consuming device state of charge $Q.SoC$ is lower than the zero state of charge $Q.SoC_0$ at the moment $CES_j.st$. $Q.SoC = \{~Q.SoC_I, ~Q.SoC(t): t \in T~\}$
where $Q.SoC_I$ is the initial state of charge of the consuming device. $Q.SoC(t)$ is the state of charge of the consuming device at moment $t$. If the energy consuming device can function properly until $CES_j$ starts e.g., $Q.SoC(CES_j.st)>Q.SoC_0$, energy services $CES_i $ and $CES_j $ can be selected together for a possible composition.    
\begin{figure}[!t]
    \centering
    \includegraphics[width=0.37\textwidth,height=4.5cm]{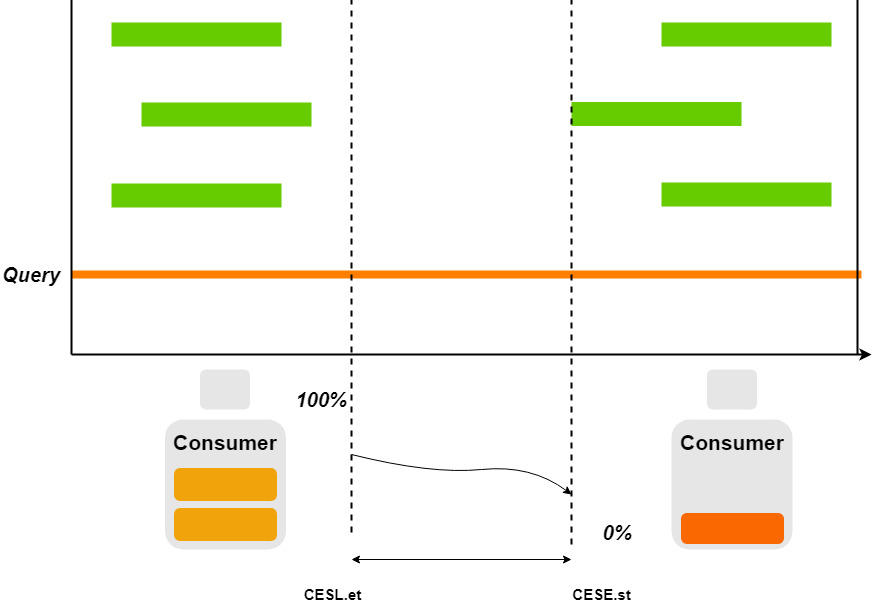}
         \caption{\small{Dividing a query based on the waiting time}}
        \label{fig:ExampleWaitintime}
    \vspace{-0.35 cm}
\end{figure}

If the consuming IoT device has enough energy to sustain until the second service, the two services are not composable. For example, in Figure \ref{fig:ExampleWaitintime} the candidate services are in two \textit{different clusters}. There is a time gap between the two clusters. The consuming device can sustain along this period of time. Hence, the energy query can be divided into two \textit{shorter queries} for more convenience of the IoT user. Moreover, the first query on the left may be dropped since the consumer has enough energy.
\vspace{-0.35 cm}
\subsection{Greedy composition of CES}
IoT users have different preferences for their energy request. An IoT user requires a certain amount of energy $Q.RE$ for a defined query time interval $[Q.t, Q.t+d]$. The user may prefer to recharge at the earliest possible time. They might also prefer to get $Q.RE$ in the shortest period of time. Another preference is to get the maximum of energy during $Q.d$. \textit{We generate different rankings for energy services based on the different user preferences e.g., earliest time, shortest time, or a maximum of energy.} If the energy query can be satisfied by one service, the best service is selected based on the \textit{user preference}. If an energy request can not be satisfied with one single service, multiple services need to be composed. 

Note that, the filtered energy service candidates are spatio-temporally composable. Energy services might be composed sequentially if they do not overlap. Services can also be composed simultaneously if they overlap during the query time interval. The intensity compatibility rule must be verified while composition. The assumption of this composition technique is that all energy services within the query duration cannot be decomposed into services of shorter intervals. The greedy composition relies mainly on selecting and composing the best services in terms of the user preferences e.g., maximum energy, earliest service, or shortest service time. 

The recursive algorithm \ref{Greedy} presents the systematic selection of the optimal set of services among the spatio-temporally composable services. The selected service $BestS$ is verified if it is composable according to the intensity compatibility rule with the already selected services $CompositeS$ (Algorithm \ref{Greedy}-Lines 5-9). If the service $BestS$ is composable, the service is added to the set of Composite service $CompositeS$ (Algorithm \ref{Greedy}-Lines 10,11). The greedy algorithm stops composing services when the user query is satisfied or no service is available for further consideration.     

\begin{algorithm}[!t]
\footnotesize
	\caption{Greedy Composition of Energy Services}
	\label{Greedy}
	\begin{algorithmic}[1]
	\State \textbf{Input: }$ Q.l$, $Q.t$ , $Q.d$, $NearbyS$
		\State \textbf{Output: }$Composite$
		\State \text{$Composite$ component energy services  during $Q.d$}
		\State $Composite \gets \emptyset$
		\If {$NearbyS \neq \emptyset$}
		\State $Composite \gets Composite \cup \emptyset$
		\State  \textbf{else}
		\State $Smax \gets Max(NearbyS) ${~~~~~~~~~~~// Select the best service among the available services}
		\State $NearbyS \gets NearbyS~-~\{Smax\} ${~~~~~~~~~~~// Update Nearby by the remaining services}
		\If {$Composable(Composite, Smax)$} {~~~~~~~~~~~// Check if the $Smax$ is composable with the previously selected services }
		\State $Composite \gets Composite \cup \{Smax\}$
		\State  \textbf{else}
		\State $Composite \gets Greedy(Q.l~, ~Q.t ~, ~Q.d~,NearbyS)$
		\EndIf
		\EndIf
		\State \Return $ Composite $
	\end{algorithmic}
\end{algorithm}
\vspace{-0.35 cm}
\subsection{Multiple local knapsack-based CES Composition}

One of our assumption is that energy services can be consumed partially. The user may switch to other energy services within the query duration. We define all the possible timestamps where the user may need to switch to better services in terms of the provided energy. Each timestamp is either the start time or the end time of available services. Simply, the user may switch to a newly available service if the new service is better than the current service. If no better service is found, at the end of a service, the user needs to switch to the best available service. We divide the query duration into several time slots based on these timestamps (see vertical lines in Figure ~\ref{fig:Examplechunking}). The time slots represent the arrival time of a new service or the exit time of an existing service (Algorithm \ref{FKP}-Lines 5-13). For example, the start time of $CES3$ defines the first time slot. The start time of $CES4$ defines the second time slot. After dividing the query duration based on the defined timestamps, some chunks may have a very short period of time which might not respect the \textit{composition eligibility rule}. Energy consumers may loose energy more than the provided energy by the selected service within this thin chunk due to the connection establishment.
We propose a \textit{smoothing technique} for the thin chunks in three situations as follows:
\begin{itemize}
    \item The thin chunk is created by two consecutive end times $et$ of services. We propose to eliminate the latest timestamp defined by $et$ and widen the next chunk on the right.
    \item If the thin chunk is created by two consecutive start time $st$ of services, we eliminate the earliest timestamp defined by $st$ and widen the previous chunk on the left.
    \item If the thin chunk is created by two timestamps defined by start time $st$ and end time $et$ of services, we eliminate the timestamp defined by the delimiter e.g., $st$ or $et$ of the service having the lowest value of intensity $I$.    
\end{itemize}

\begin{figure}[!t]
    \centering
    \includegraphics[width=0.35\textwidth]{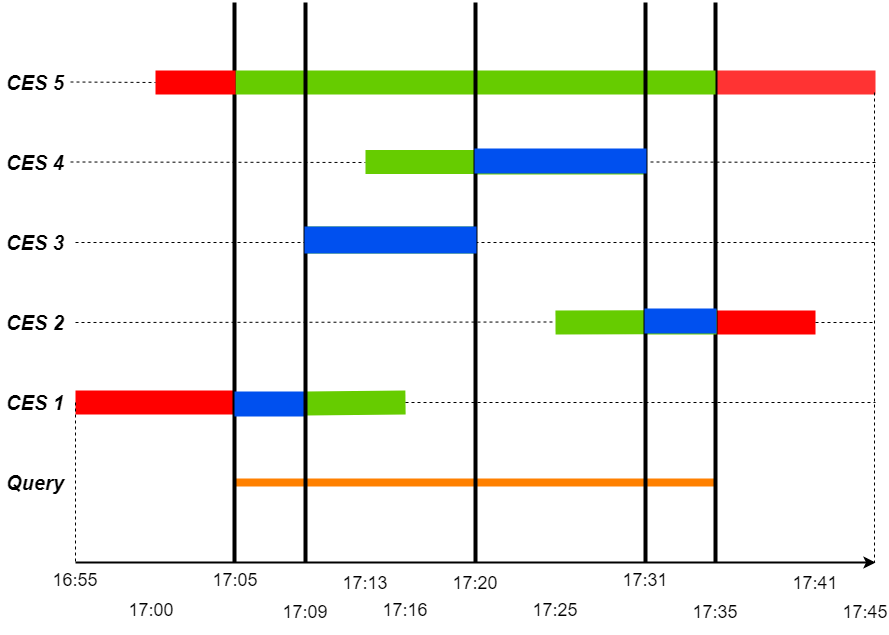}
        \caption{\small{Reducing Chunks}}
        \label{fig:reducedchunks}
    \vspace{-0.35 cm}
\end{figure}

We propose a multiple knapsack composition approach which aims to select an optimal set of partial energy services based on their deliverable energy capacity $DEC$ and the defined chunks. Partial energy services which are available within the chunk are constrained by the \textit{intensity compatibility rule}. The sum of the intensity of the simultaneously composed partial services cannot exceed the compatible intensity of the consuming device $Q.I$. In this situation, we have a local partial services composition for each chunk. The resulting composite services from all chunks are grouped into a global composite service which provides the maximum amount of energy from all the composable service candidates (Algorithm \ref{FKP}-Lines 15-27). The composition  problem in each chunk is formulated as a $0/1$ knapsack problem. A knapsack problem is the selection of a set of items having weights and values by maximizing the total value of selected items considering the limited weight capacity of the knapsack (Algorithm \ref{FKP}-Lines 19-25). We interpret each partial energy service within the chunk as an item. The service intensity $I$ is considered as a weight of the partial services. $DEC$ is  considered as a value. The knapsack weight capacity is defined by the maximum capacity of the intensity compatible service.  

\begin{algorithm} [!t]
\footnotesize
	\caption{Multiple Knapsack composition}\label{FKP}
	\begin{algorithmic}[1]
	\State \textbf{Input: }$ Q.l$, $Q.t$ , $Q.d$, $NearbyS$
		\State \textbf{Output: }$Composite$ 
		\State \text{$Composite$ component energy services  during $Q.d$}
		\Statex \text{ // Chunking the query duration}
		\State $Chunk_{0}.st \gets Q.t $
		\State \textbf{For} \text{$ int~t= Q.t~to~ Q.t+d $} \textbf{do}
		\If {\text{$(\forall~CES\in NearbyS~and~t = CES.st ~or~t = CES.et)$}}
		\State $ Chunk_{i}.et\gets t $
		\EndIf
		\Statex \text{ // create new chunk}
		\If {$t \neq Q.t+d  $}  
		\State $ Chunk_{i+1}.st\gets t $
		\State $ i \gets i+1 $
		\EndIf
		\State \textbf{End For}
		\State \text{Smooth thin chunks}
		\Statex \text{// apply 0/1 knapsack optimization at each chunk}
		\State \textbf{For each chunk}
		\Statex \text{// miniComposite is the local composition in a chunk}
		\Statex \text{// miniCES is the set of  partial service within a chunk}
		\State $miniComposite \gets \emptyset$
		\State \textbf{While}\text{($miniCES \neq \emptyset$)}
		\State    \text{$Smax  \gets Max( ~miniCES )$} 
		\State    $miniCES  \gets miniCES~-~\{Smax\}$ 
 		\If        {($ Composable(minicomposite~,~ max) $ ) }
 		\State    \text{$miniComposite \gets miniComposite \cup \{max\}$}
 		\EndIf
 		\State \textbf{End While}
 		\State \text{$Composite \gets Composite \cup \{ miniComposite\}$}
 		\State \textbf{End For}
		\State \Return $ Composite $
	\end{algorithmic}
\end{algorithm}

\begin{table*}
	\begin{center}
		\caption{Fractional service composition $Q$}
		\label{tab:table3}
		\begin{tabular}{|c||c |c| c| c| c| c| c| c||c|} 
			\hline
			\textbf{Service}&\textbf{$ I~.~Tsr~.~\alpha $}& \textbf{Slot1} & \textbf{Slot2}& \textbf{Slot3}& \textbf{Slot4}& \textbf{Slot5}& \textbf{Slot6}& \textbf{Slot7}& \textbf{Total $ DEC $}\\
			\hline
			$CES_{1}$ & 0.92&\textbf{61.33}& 61.33 & 47.34 & ~-~ & ~-~ & ~-~ & ~-~ & 170\\
			$CES_{2}$ & 0.78 & ~-~& ~-~&~-~ & ~-~& ~-~ & 78.00 & \textbf{52.00} & 130\\
			$CES_{3}$ & \textbf{1.09} &~-~&\textbf{72.67}& \textbf{54.66} & \textbf{72.67} & ~-~ & ~-~ &~-~ & 200\\
			$CES_{4}$ & 0.90 & ~-~&~-~& 45.00 & 60.00 & \textbf{75.00} & \textbf{90.00} & ~-~ & 270\\
			$CES_{5}$ & 0.66 & 44.00&44.00& 33.00 & 44.00 & 55.00 & 66.00 & 44.00 & 330\\
			\hline
		\end{tabular}
	\end{center}
\end{table*}
\vspace{-0.35 cm}
\subsection{Heuristic-based spatio-temporal CES composition}

Our objective is to improve the multiple local knapsack based composition technique. Solving a $0/1$ knapsack algorithm at each chunk provides an exhaustive exploration of all the possible compositions which has \textit{exponential runtime efficiency}. The heuristic aims to merge some chunks to reduce the number of local optimizations. An approximate way to preserve the optimal compositions is to merge two consecutive chunks having the same service providing the maximum energy. Most probably, this latter service is selected by the $0/1$ knapsack algorithm in addition to other composable services. The heuristic selects only two miniservices, the first miniservice providing the maximum energy amount and the next one providing the maximum amount of energy at each chunk. The selected miniservices have to be composable in terms of compatibility. If there is no composable service with the service providing the maximum energy amount at that chunk. the heuristic takes only this latter.

First, we eliminate all the delimiters where the service providing maximum energy does not change over two consecutive chunks (Algorithm \ref{heuristic}-Lines 15-21). A $0/1$ knapsack algorithm is applied then for all the new chunks (Lines 22-31). Table \ref{tab:table3} presents the provided energy by each service at each chunk after chunking the query duration (see figure \ref{fig:Examplechunking}). Chunks 2, 3, and 4 are merged together because the service $CES3$ provides the maximum amount of energy over these three chunks. Time chunks 5 and 6 are also merged because the service $CES4$ keeps providing the  the maximum amount of energy along these two chunks. In the illustrated example, the number of initial chunks has been reduced from 7 slots to 4 slots (see figure \ref{fig:reducedchunks}). 
The number of new chunks $N_{nc}$ may decrease slightly or does not change  compared to the number of the original chunks $N_{orc}$. $N_{nc}$ may also decrease significantly. The number of the original chunks $N_{orc}$ can be  in a larger or smaller scale compared to $N_{nc}$. The intuitive explanation of the steep decrease in the number of new chunks $N_{nc}$ is the existence of a small number of services with high intensity along the query duration. Conversely, if $N_{nc}$ is slightly lower than $N_{orc}$, it means that all the available services along the query duration are short services in terms of availability  and comparable in terms of the provided energy.

\begin{algorithm}[!t]
\footnotesize
	\caption{Heuristic based space reduction}\label{heuristic}
	\begin{algorithmic}[1]
	    \State \textbf{Input:} $Q.l$, $Q.t$ , $Q.d$, $NearbyS$
		\State \textbf{Output: }$Composite$
		\State \text{$Composite$ component energy services during $Q.d$}
		\Statex \text{ // Chunking the query duration}
		\State $Slot_{0}.st \gets Q.t $
		\State \textbf{For} \text{$ int~t= Q.t~to~ Q.t+d $} \textbf{do}
		\If {\text{$(\forall~CES\in NearbyS~and~t = CES.st ~OR~t = CES.et)$}}
		\State $ chunk_{i}.et\gets t $
		\EndIf
		\Statex \text{ // create new chunk}
		\If {$t \neq Q.t+d  $}  
		\State $ chunk_{i+1}.st\gets t $
		\State $ i \gets i+1 $
		\EndIf
		\State \textbf{End For}
		\State \text{Smooth thin chunks}
		\Statex \text{ // Update chunking by maximum services}
		\State \textbf{For all Chunks}
		\If        {($ Max(Chunk_i) = Max(Chunk_{i+1}) $ ) }
		\State $Chunks_i.et  \gets Slot_{i+1}.et$
		\State \text{Delete ($Chunk_{i+1}$)}
		\EndIf
		\State \textbf{End For}
		\State \text{Sort ($New Chunk ~array$)}
		\Statex \text{// miniComposite is the local composition in a chunk}
		\Statex \text{// miniCES is the set of  partial service within a chunk}
		\State $miniComposite \gets \emptyset$
		\State \textbf{While}\text{($miniCES \neq \emptyset$) and ($ miniComposite \leq 2$)}
		\State    \text{$Smax  \gets Max( ~miniCES )$} 
		\State    $miniCES  \gets miniCES~-~\{Smax\}$ 
		\If        {($ Composable(minicomposite~,~ max) $ ) }
		\State    $miniComposite \gets miniComposite \cup \{max\}$
		\EndIf
		\State \textbf{End While}
		\State \text{$Composite \gets Composite \cup \{ miniComposite\}$}
		\State \textbf{End For}
		\State \Return $ Composite $
	\end{algorithmic}
\end{algorithm}

The complexity of the three proposed algorithms can be estimated based on the number of available services and the number of chunks. Since the greedy algorithm is short sighted, it has an efficient runtime complexity with a limited performance. The runtime complexity of the greedy composition can be summarized into the systematic selection of $n$ available services multiplied by $n-1$ composability control with that selected service $\mathcal{O}(n^2)$. The complexity of multiple knapsack-based composition can be estimated based on the number of chunks $C$ and the complexity of the algorithm solving 0/1 knapsack problem at each chunk with respect to the intensity compatibility. We consider a dynamic programming solution for the 0/1 knapsack problem. The runtime complexity of the multiple knapsack-based composition is $\mathcal{O}(Cm^2)$. If we consider $m$ as the number of available partial services within a chunk. The difference between the heuristic-based composition and the multiple knapsack-based composition is the significant reduction in the number of chunks of the energy query and considering only two partial services at each chunk. The heuristic-based composition complexity becomes $\mathcal{O}(Cm)$.

\vspace{-0.45cm}
\section{Experiment results}
\vspace{-0.15 cm}

\begin{table*}[!t]
\centering
\small
\caption{Parameters of the experiments setting}
\begin{tabular}{|l|c|c||l|c|c|}
\hline
\multicolumn{3}{|c||}{\footnotesize{Crowdsourced energy service}}                                                                                                          & \multicolumn{3}{c|}{\footnotesize{Energy query}}                                                                                                                                                                                  \\ \hline
\multicolumn{1}{|c|}{\footnotesize{QoS}} & \footnotesize{Dataset}                                                              & \footnotesize{value}                                                       & \multicolumn{1}{c|}{\begin{tabular}[c]{@{}c@{}}\footnotesize{Query parameters}\end{tabular}} & \footnotesize{Dataset}                                                              & \footnotesize{value}                                                       \\ \hline
\footnotesize{Start time}                & \footnotesize{Yelp}                                                                 & \footnotesize{Check-in}                                                    & \footnotesize{Start time}                                                                        & \footnotesize{Yelp}                                                                 & \footnotesize{Check-in}                                                    \\ \hline
\footnotesize{End time}                  & \footnotesize{Uniform distribution}                                                               & \footnotesize{Uniform distribution}                                                      & \footnotesize{End time}                                                                     & \begin{tabular}[c]{@{}c@{}}\footnotesize{Uniform distribution}\end{tabular}              & \begin{tabular}[c]{@{}c@{}}\footnotesize{Uniform distribution}\end{tabular}     \\ \hline
\footnotesize{Energy capacity}           & \begin{tabular}[c]{@{}c@{}}\footnotesize{Renewable energy sharing}\end{tabular} & \begin{tabular}[c]{@{}c@{}}\footnotesize{Provided energy}\end{tabular} & \footnotesize{Energy capacity}                                                                   & \begin{tabular}[c]{@{}c@{}}\footnotesize{Renewable energy sharing}\end{tabular} & \begin{tabular}[c]{@{}c@{}}\footnotesize{Consumed  energy}\end{tabular} \\ 
\hline
\end{tabular}
\label{tab:simparam}
\end{table*}


We compare the proposed composition algorithms with two different composition algorithms, priority-based resource scheduling algorithm \cite{ghanbari2012priority}, a QoS-aware time constrained composition algorithm \cite{ai2011qos}. We consider energy services as resources and maximizing energy as a priority for the priority-based resource scheduling algorithm. This algorithm relies mainly on selecting and composing the best energy services without chunking them.  The QoS-aware service composition transforms the composition into a time constrained optimization problem then solve it using genetic algorithms. After chunking the query duration, the algorithm finds the optimal composition which maximizes the provided energy within the query duration.   

We evaluate the different composition techniques by two sets of experiments. First, we evaluate the \textit{effectiveness} and \textit{feasibility} of each composition technique in terms of the number of successfully served queries. We compare how close the {\em completeness} (i.e., the ratio of successfully served queries over the number of all queries) given by our heuristic-based composition algorithm to the completeness of the optimal composition
given by the  proposed brute-force-like algorithm (Multiple knapsack-based composition algorithm) \cite{mabrouk2009qos}, \cite{fellows2012local}. Second, we evaluate the \textit{scalability} of each composition algorithm by measuring the computation time while varying the number of energy services. Energy services have different spatio-temporal features and provided energy amounts. The experiments only test the performance of the composition framework from a single consumer perspective to evaluate the runtime efficiency and the effectiveness of the proposed algorithms in different scenarios. The effectiveness test hypothesis is “The more available energy services in a confined area (e.g., coffee shop), more successfully provisioned energy queries by the proposed composition framework”. In the future work, we will extend the composition framework to process multiple parallel energy queries by labeling the already reserved services using different priority strategies. 
\vspace{-0.35cm}
\subsection{Datasets and experiment scenarios}

\begin{table}[!t]
\centering
\caption{Statistics of the crowdsourced IoT environment}
\begin{tabular}{|l|l|}
\hline
\footnotesize{Parameter} & \footnotesize{Range of values} \\
\hline
\footnotesize{Confined areas}                & \footnotesize{8280}                                \\ \hline
\footnotesize{Queries}                       & \footnotesize{5000}                                \\ \hline
\footnotesize{Services}                      & \footnotesize{5000-50 000}                         \\ \hline
\footnotesize{Duration of a service}         & \footnotesize{10-60  minutes}                      \\ \hline
\footnotesize{Duration of a query}           & \footnotesize{5-120 minutes}                      \\ \hline
\footnotesize{Provided energy}               & \footnotesize{50-1000 mAh}                        \\ \hline
\footnotesize{Energy requirement}            & \footnotesize{100-800 mAh}                       \\ \hline
\end{tabular}
\label{tab:simStat}
\end{table}

We create a scenario of crowdsourced IoT environment close to the reality to evaluate the performance of the proposed composition approach.

To the best of our knowledge, there is no dataset about wireless energy sharing among human-centric IoT devices. We consider that crowdsourced energy services are provided from wearables or the spare energy of the smartphone batteries of IoT users.  We create an energy crowdsourcing environment close to reality based on a renewable energy sharing environment \footnote{https://data.gov.au/dataset}. A set of 25 houses daily harvest, consume and provide energy from their solar panels for two years [April 2012 to March 2014]. They are considered either as energy providers or consumers. Energy consumption and production is recorded every 30 minutes. Each house has 730 daily records ($365$x$2$). Each daily record has $48$x$2$ fields for the produced and the consumed energy for each day. 

The crowdsourced energy service QoS parameters are defined based on these records. We normalize all the energy measurement values for all records from {\em Watt hour to miliampere hour (mAh)} to mimic the energy provided and consumed by IoT devices e.g., smartphone and wearables. The deliverable energy capacity $DEC$ of IoT energy services $S_i$ starting at $S_i.st$ and ending at $S_i.et$  is defined by {\em randomly} matching a daily record of a provider from the renewable energy sharing environment considering only the energy produced during the same period of time. Similarly, the energy requirement of a query $Q.RE$ is also generated and normalized from the daily energy consumption of the houses according to the query duration $Q.du$. 

We use {\em Yelp}\footnote{https://www.yelp.com/dataset} dataset to define the {\em spatio-temporal features} of IoT energy services and queries by {\em check-in} and {\em check-out} timestamps of people to a confined area. The dataset contains people's check-ins information into different confined areas e.g., coffee shops, restaurants, libraries etc. We consider these people as IoT users who are either energy providers or consumers. For example, the start time $st$ of an energy service provided by an IoT user is the time of their check-ins into a coffee shop. Energy queries time $Q.t_s$ and duration $Q.du$ are also generated from check-in and check-out timestamps of consumers. We use a {\em uniform distribution} to augment the check-in data and generate different {\em durations} for energy services and queries. 

We define different scenarios to set the quality parameters of IoT based energy services. The scenarios are defined by the {\em capacity} of the provided energy and the availability {\em duration} of services. The service duration varies from 5 minutes to 1 hour to cover different scales of IoT devices. We also define multiple scenarios of energy queries. Energy queries are ranging from \textit{short duration queries} with a low amount of required energy capacity to queries with \textit{long duration} and high amount of required energy. The query duration $Q.d$ varies between 10 minutes and 2 hours. Tables \ref{tab:simparam} and \ref{tab:simStat} present statistics about the used datasets. 
\begin{figure}[!t]
    	\centering
		\includegraphics[width=0.34\textwidth]{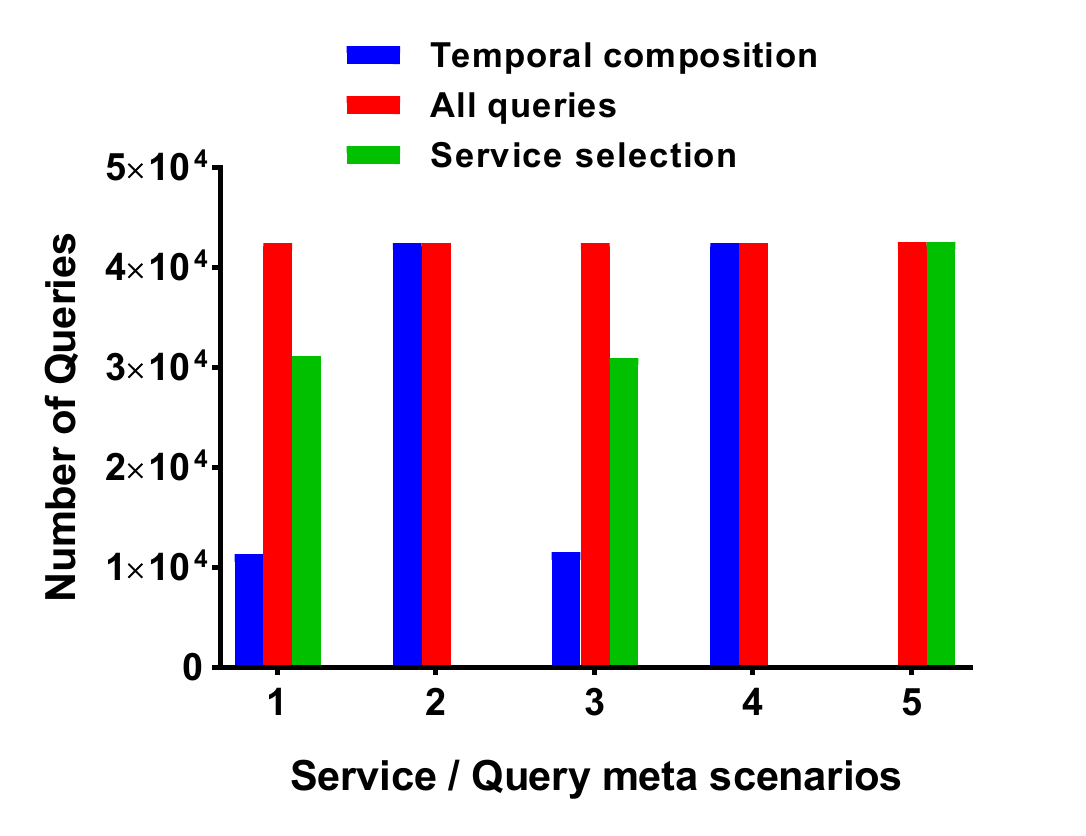}
        \caption{\small{Meta scenarios of temporal composition}}
        \label{fig:metascen}
        \vspace{-0.45cm}
\end{figure}
\begin{figure*}[t!]
 \centerline{
  \subfloat[]{\label{fig:short_Complet}\includegraphics[width=.3\textwidth]{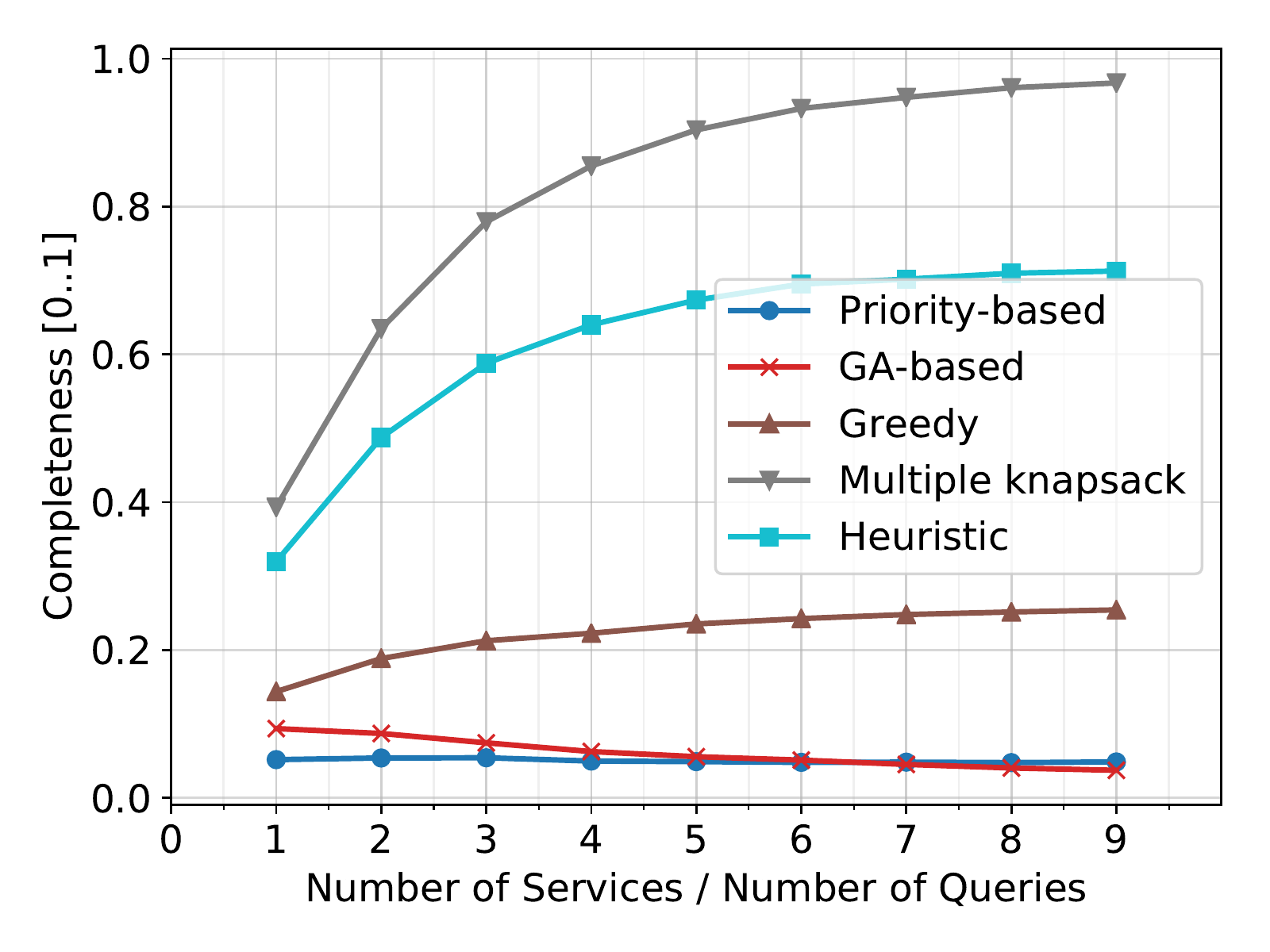}}
  \hfill
  \subfloat[]{\label{fig:long_Complet}\includegraphics[width=.3\textwidth]{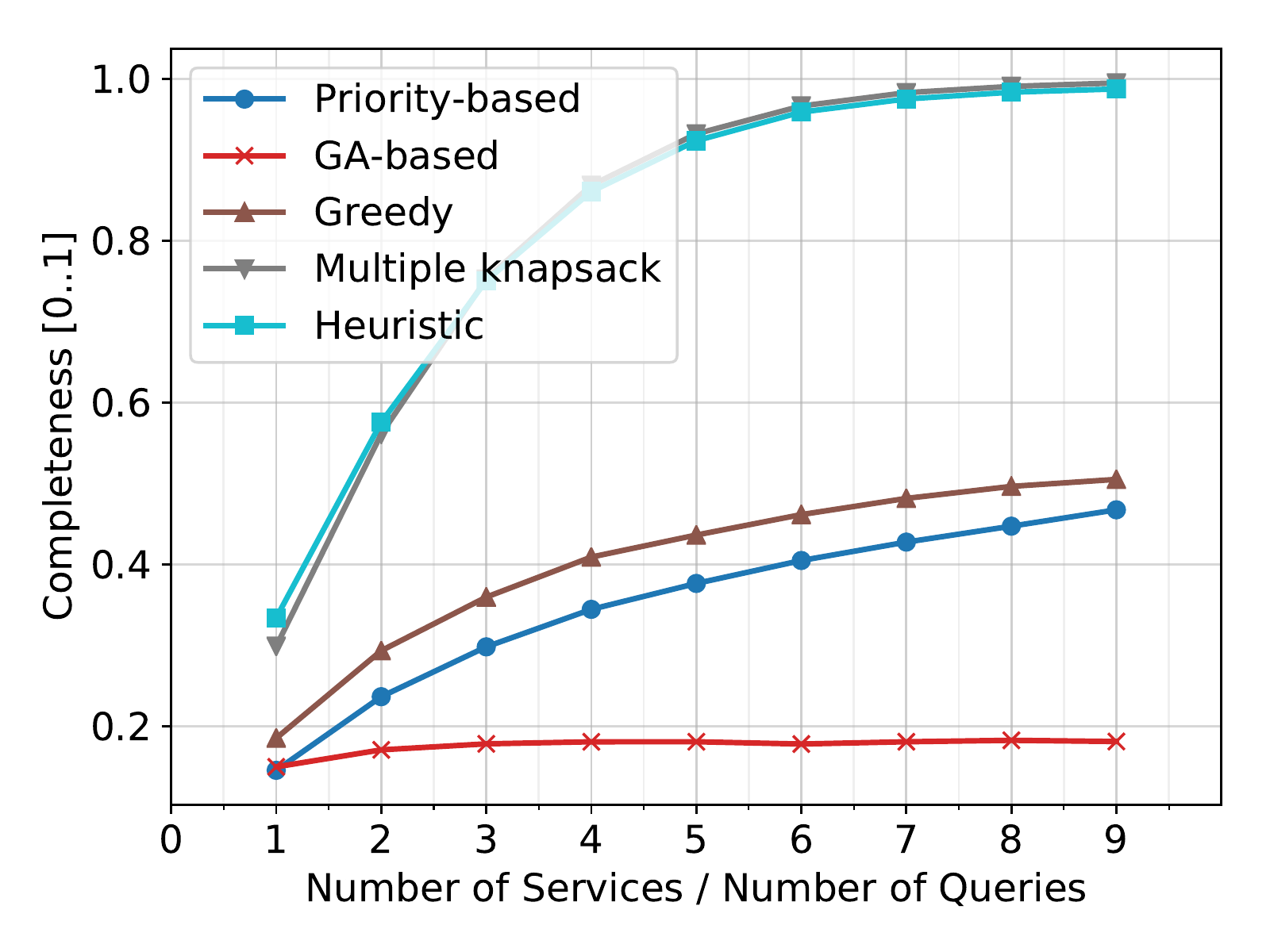}}
  \hfill
  \subfloat[]{\label{fig:agg_Complet}\includegraphics[width=.3\textwidth]{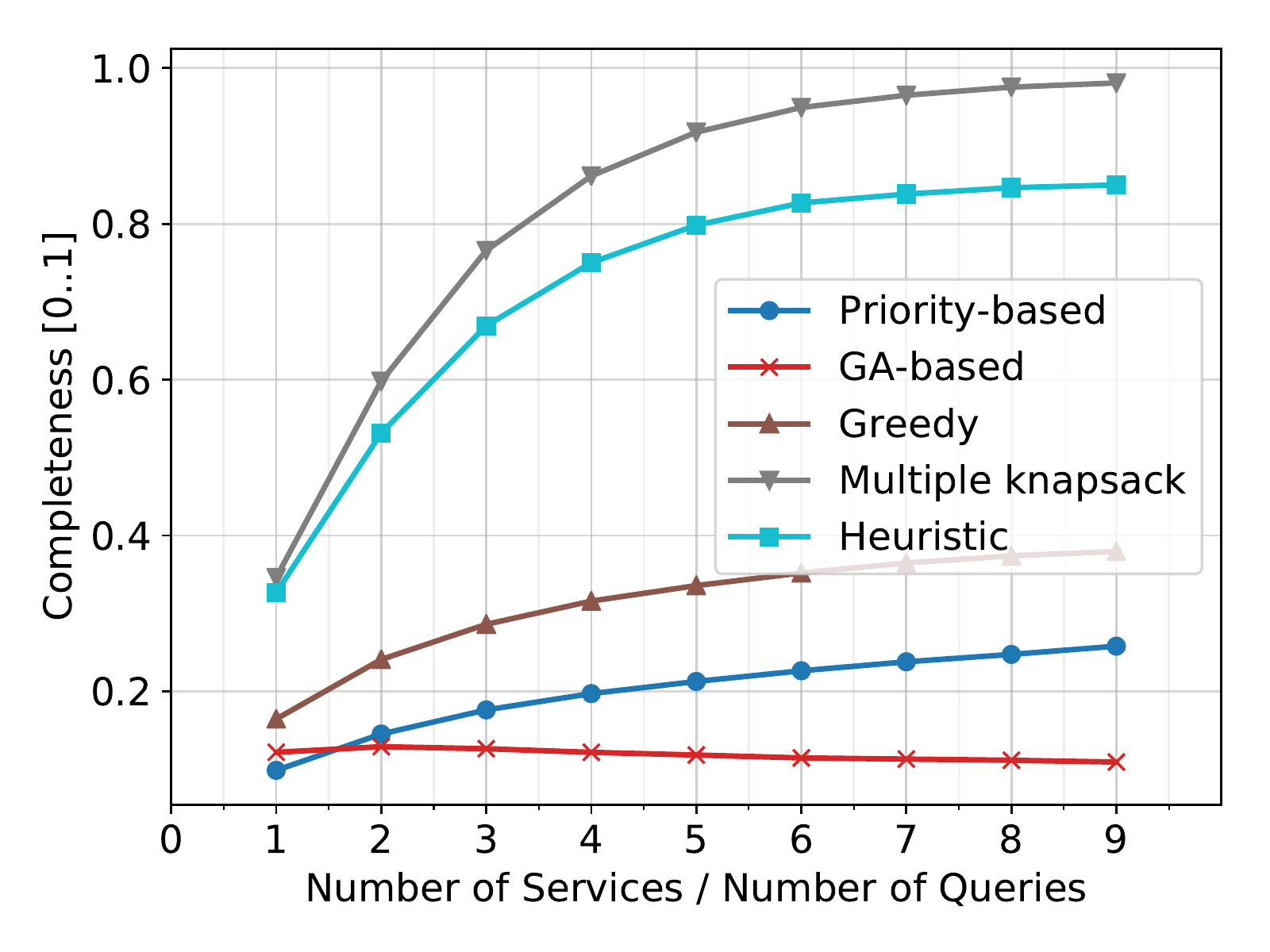}}
  }
  \caption{\small{Successfully served queries Vs number of services in (a) Short services, (b) Long services, (c) all services} }
        \label{fig:Complet}
        \vspace{-.5cm}
\end{figure*}
\begin{figure*}[t!]
 \centerline{
  \subfloat[]{\label{fig:Gplot}\includegraphics[width=.32\textwidth]{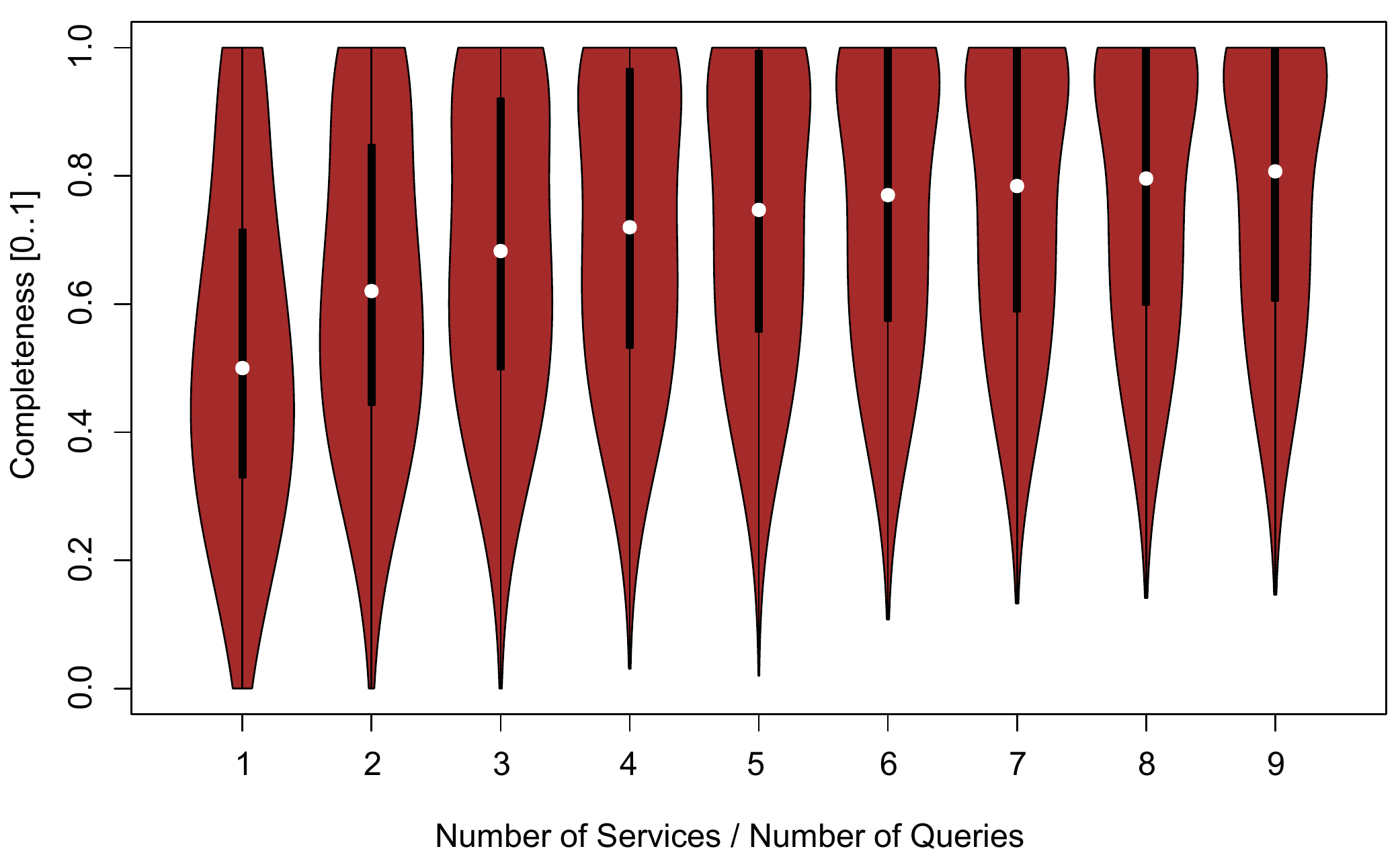}}
  \hfill
  \subfloat[]{\label{fig:HPlot}\includegraphics[width=.32\textwidth]{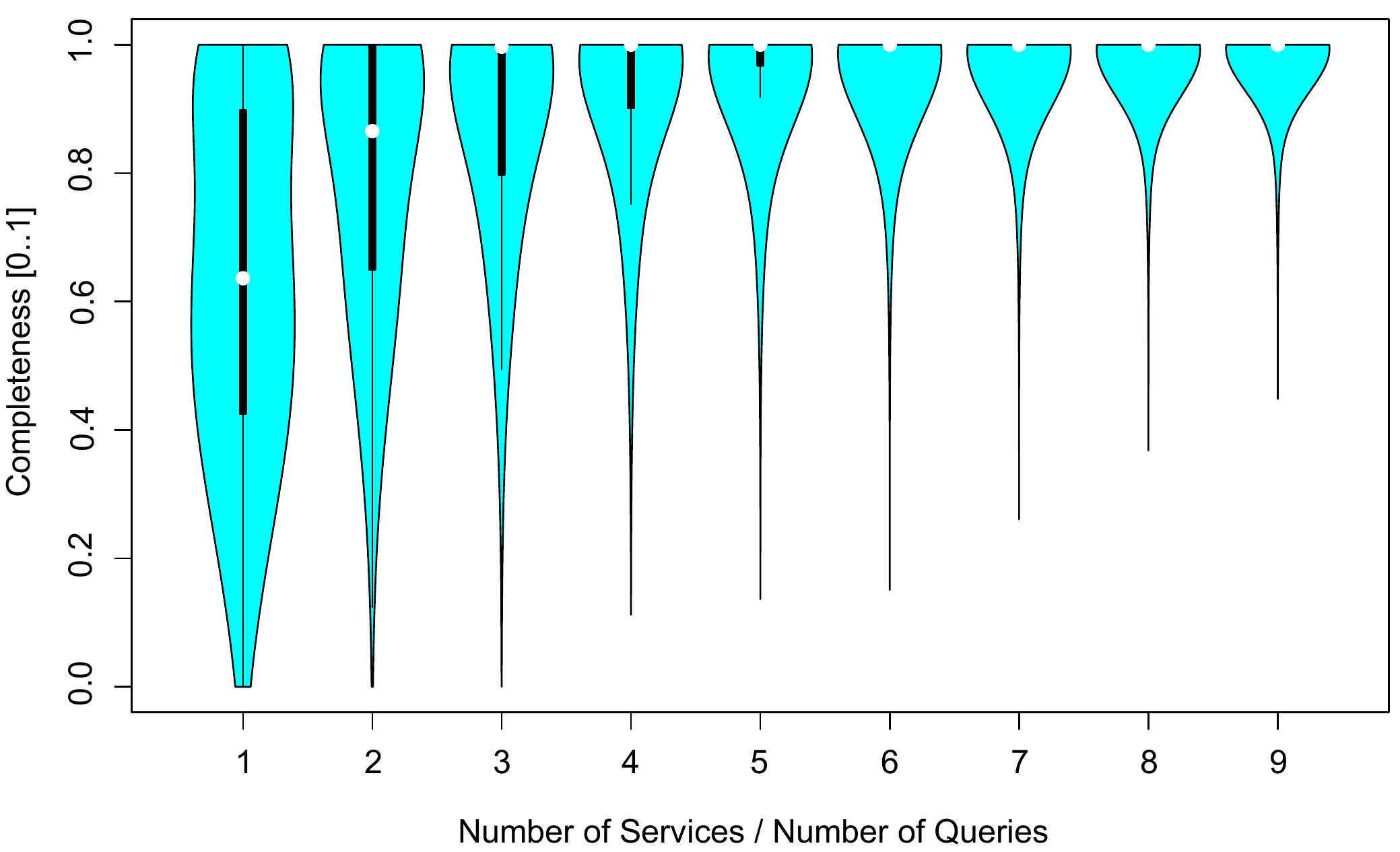}}
  \hfill
  \subfloat[]{\label{fig:MPlot}\includegraphics[width=.32\textwidth]{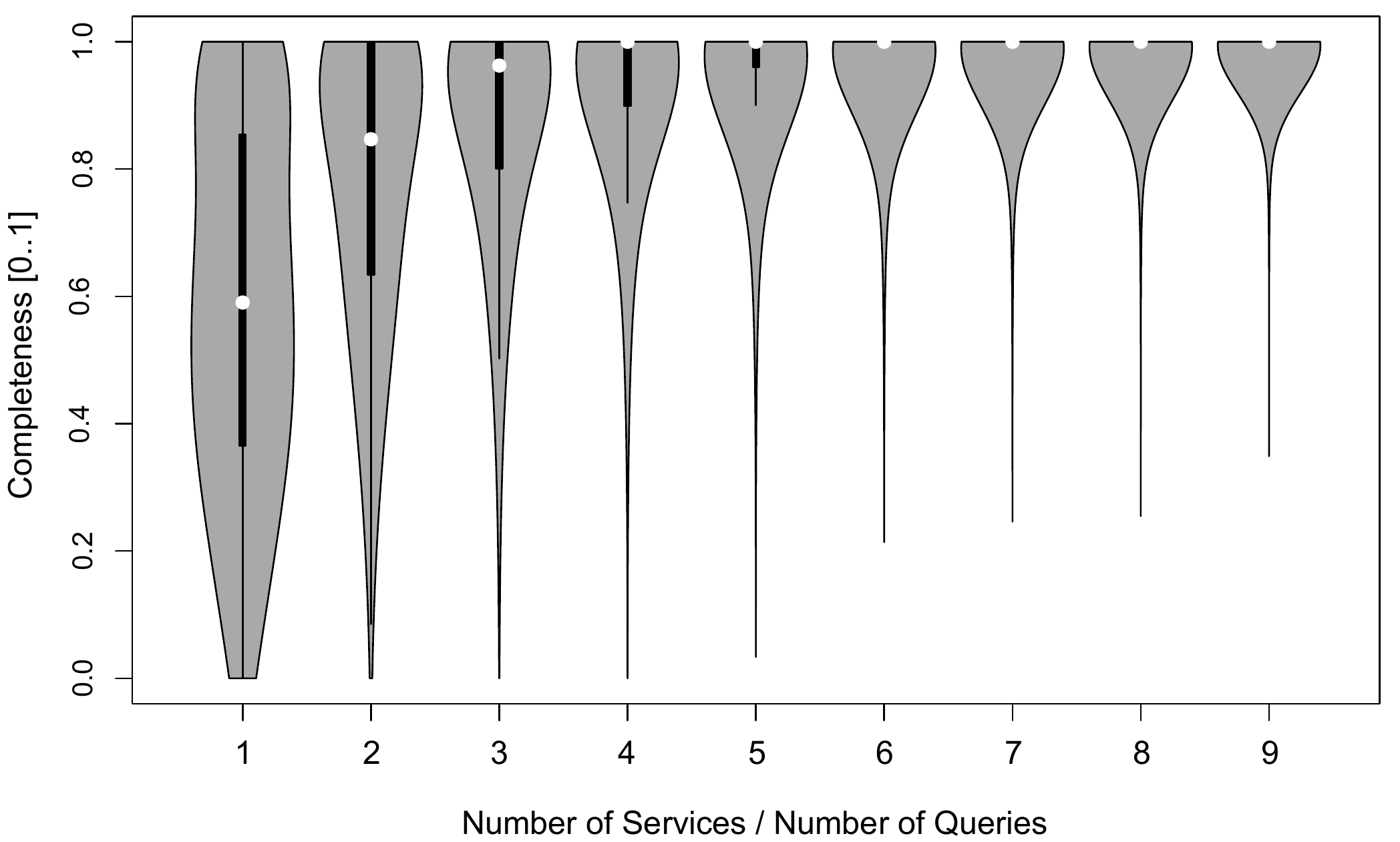}}
  }
  \caption{\small{Distributions of successfully served queries Vs number of services in (a) Greedy composition, (b) heuristic-based composition, (c) multiple knapsack-based composition}}
        \label{fig:Boxplots}
        \vspace{-.5cm}
\end{figure*}

We grouped scenarios requiring composition into 5 different meta scenarios. Two meta scenarios $(Meta Scen~1,~ Meta Scen ~2)$  defined by services with {\em short duration} delivering energy to queries with short and long duration. The next two meta scenarios $(Meta Scen~3,~ Meta Scen ~4)$ are defined by  services with {\em long duration} delivering also energy to queries with short and long duration. The last Meta scenario $(Meta Scen~5)$ represents an aggregated view of all the composition scenarios. We consider another energy quality attribute to verify the intensity compatible composability rule. We modify the intensity of each energy service in the range $CES.I \in [0.5~,~1.5~A]$ considering the variety of human centric IoT devices. We also modify the compatible intensity of each energy query within the interval $Q.I \in [1~,~2.5~A]$.

We investigate the effectiveness and the computation time of the three proposed composition techniques by a large number of energy services. 50,000 different IoT users have been identified from Yelp dataset which includes multiple check-ins into multiple confined areas. We modify the ratio between the number of queries and the number of services ($N.CES/N.Q$) among the IoT users from $1$ (the number of services is equal to the number of the queries) to $9$ (services are nine times the number of queries). In this paper, we only focus on the composition for single consumer's perspective. The scalability is reflected by the run-time efficiency of the composition algorithms. The coordination of parallel energy queries will be considered in the future.
\vspace{-0.4 cm}
\subsection{Effectiveness}

We investigate the effectiveness of the proposed composition techniques by measuring the \textit{completeness} of energy queries versus the ratio $N.CES/N.Q$. The ratio shows the number of services compared to the number of queries. We define a threshold parameter $SQ$ for energy queries. We consider a query is successfully served by the neighbor services if $SQ$ from the query has been served. For example, we consider the energy query $Q_i$ is successfully served if $80\%$ from the required energy has been provided by the neighbor devices e.g., $SQ=0.8$. We vary $SQ$ from $0.7$ to $1$ for the different composition techniques to measure the ratio of the number of successful queries to the total number of queries. We consider the average value for each technique. We define completeness parameter as the number of the successfully served queries to the number of all queries. The highest value of the completeness parameter is $1.0$.

Figure  \ref{fig:Complet} shows the completeness parameter  with respect to the $N.CES/N.Q$. In terms of performance, the greedy, QoS aware (GA-based), and the priority-based composition algorithm present low completeness score compared to the heuristic-based and the multiple knapsack-based composition. The GA-based composition has the lowest completeness score of regardless the number of services and the length of their duration (see figures   \ref{fig:short_Complet}, \ref{fig:long_Complet} and \ref{fig:agg_Complet}). This performance can be explained by the inability of QoS-aware composition to consider simultaneous services and the genetic algorithm cannot check the intensity composability.   

The greedy algorithm and the priority-based composition present competitive result because these two algorithms cannot change a selected service in the midst of a composition unless this service ends. The greedy composition performs better with short services. The short duration services allow the greedy algorithm to combine a large number of services. However, the priority-based composition performs better than the greedy algorithm in long services. Committing to a selected long service will definitely make the algorithm miss other better options along the selected service (see figure  \ref{fig:long_Complet}).

Overall, the heuristic-based composition presents competitive results as the multiple knapsack based composition (see figure  \ref{fig:Complet}) especially for long services (see figure  \ref{fig:long_Complet}). Because, the new heuristic-based chunking has no big difference with the initial chunking of the query with long services. A query usually has less chunks with long services. Figure  \ref{fig:short_Complet} reflects the astonishing performance of the multiple knapsack based composition with short services. Chunking the query duration based on short duration services leads to more multiple knapsack optimizations. In contrast, The heuristic is not widely affected by the number of chunks. This latter depends on the length of the service duration. First, it defines the same number of chunks as the multiple knapsack algorithm. Chunks are widened based on services providing the maximum energy amount. Thus, the number of chunks is significantly reduced in the the heuristic-based composition approach.\\
We analyse the distribution of the successfully served queries for the proposed algorithms to ascertain the previous effectiveness evaluation. The previous effectiveness evaluation only considers the average number of successfully served queries. In this set of experiment, we vary the ratio $N.CES/N.Q$ and define the corresponding boxplot for all the proposed composition algorithms (see figure \ref{fig:Boxplots}). The greedy composition presents sparse distribution of the completeness score even with the increasing number of services (see figure \ref{fig:Gplot}). However, the multiple knapsack and the heuristic-based compositions show a similar behavior. The sparsity of the completeness scores decreases significantly when the number of available energy services increases (see figure \ref{fig:MPlot} and \ref{fig:HPlot} ). 
This can be explained by The fact that the more available nearby services, the more likely the composition will fulfill the energy requirements. In particular, when the ratio $N.CES/N.Q$ is higher than 6, almost all the compositions are successfully served.

\vspace{-0.35 cm}
\subsection{Scalability}
\vspace{-0.1cm}
\begin{figure*}[htb]
      \centerline{
  \subfloat[]{\label{fig:short_CPU}\includegraphics[width=.3\textwidth]{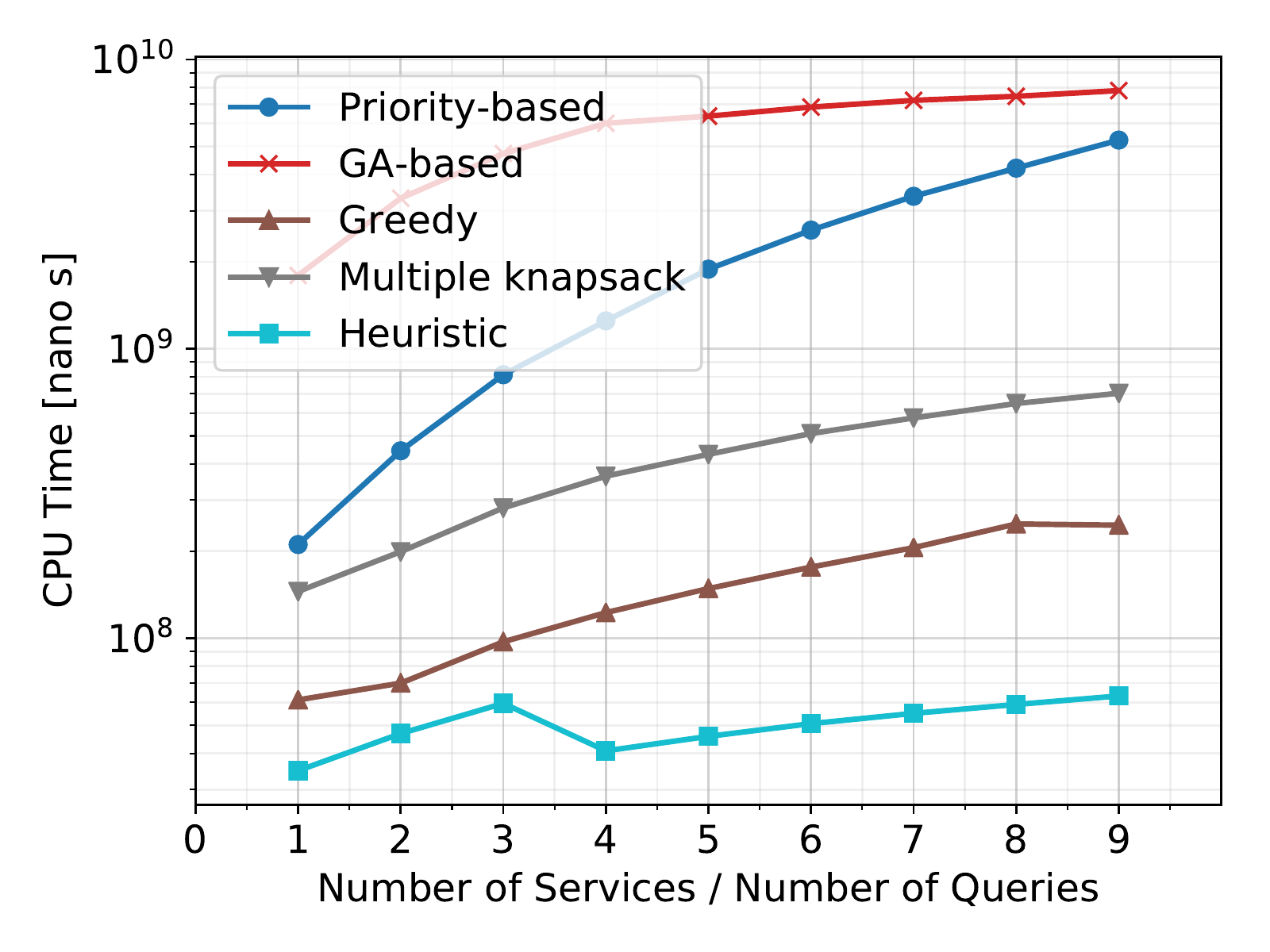}}
  \hfill
   \subfloat[]{\label{fig:long_CPU}\includegraphics[width=.3\textwidth]{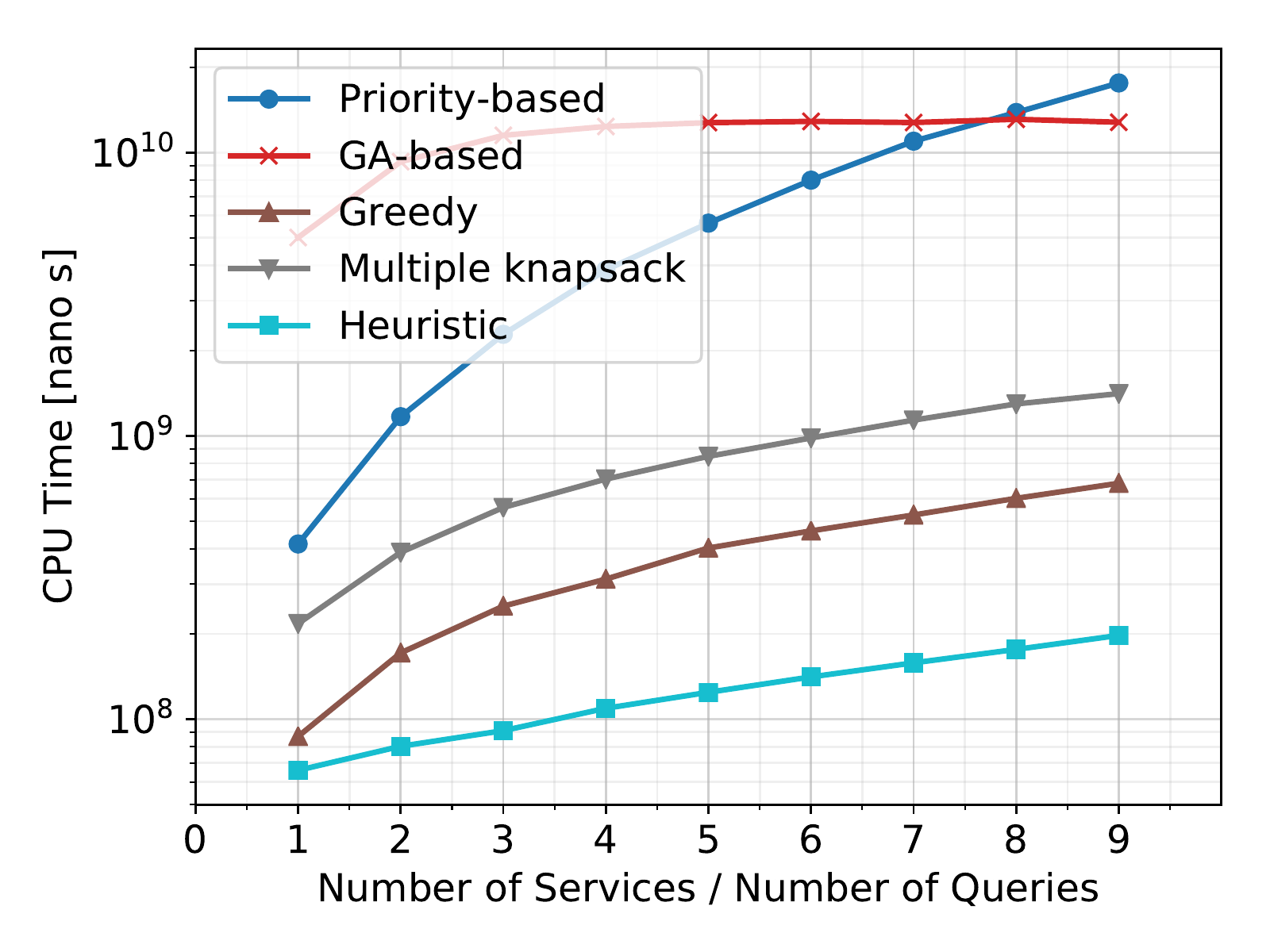}}
   \hfill
   \subfloat[]{\label{fig:agg_CPU}\includegraphics[width=.3\textwidth]{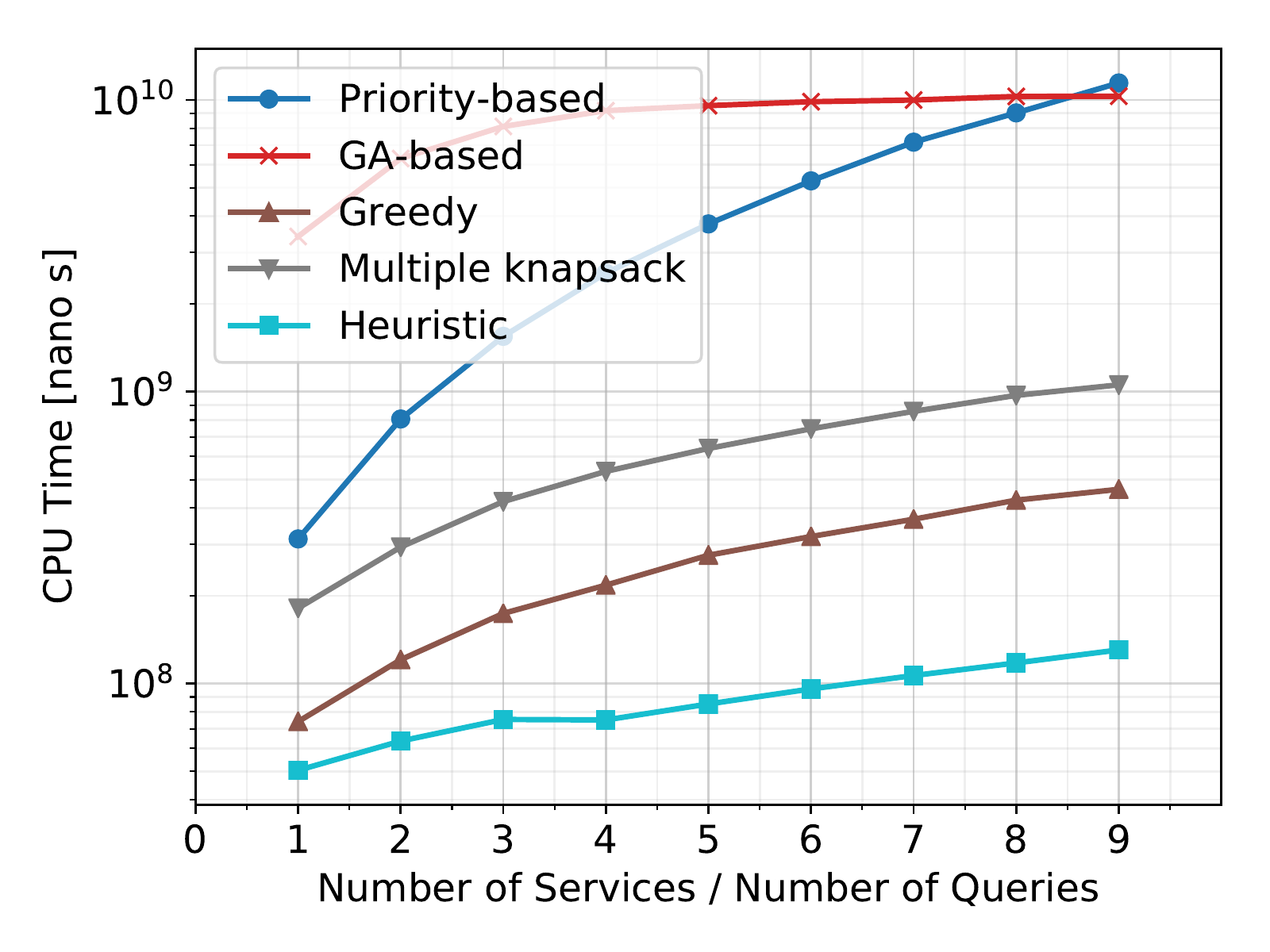}}
  }
  \caption{\small{Execution time Versus number of services in (a) Short services, (b) Long services, (c) All services} }
        \label{fig:CPU}
        \vspace{-.5cm}
\end{figure*}

We measure the average execution time of each composition technique conducted in the five different meta scenarios. Figure \ref{fig:CPU} presents the execution time of each technique with respect to the ratio $N.CES/N.Q$. We consider the number of nearby services varies from 5 to 50 energy services for each query. The results show that the execution time increases as the ratio  $N.CES/N.Q$ increases. This performance behavior is expected from all the composition techniques because of the increase in the number of services from IoT users. The heuristic-based composition technique is performing in lower execution time compared to the other compositions in all scenarios (see figures \ref{fig:short_CPU}, \ref{fig:long_CPU},). The multiple knapsack based composition is consuming more CPU time in all scenarios compared to the greedy and the heuristic-based techniques. It starts by chunking the query duration based on the presence and the end of energy services. A $0/1$ knapsack optimization then is performed at each chunk.

The number of optimization operations performed by each composition technique explains the difference between the heuristic and the multiple knapsack composition techniques. The heuristic based composition aims to minimize the number of optimizations by reducing the number of chunks. Moreover, the heuristic takes less CPU time than the greedy algorithm in all scenarios because it does not apply the $0/1$ knapsack algorithm at each chunk. Instead, it selects the mini service providing the maximum amount of energy at that chunk. It adds then the second mini service providing the amount of energy as long as it is composable in terms of compatibility.

The behavior of the greedy composition is explained by the fact that it does not perform any optimization while selecting services. However, the priority-based and the GA-based composition techniques have a very poor runtime efficiency in the crowdsourced IoT environment. Even with a small number of available services nearby a consumer, GA-based and priority-based compositions have to find all the possible combination before selecting the optimal composite energy service. The priority-based composition finds all the possible combinations without chunking the available energy services. The GA-based composition finds all the possible combinations after chunking the query duration, which explains its highest CPU time for all the scenarios.
\vspace{-0.4 cm}
\section{Related work}\label{rewo}
The background of our work comes from three different areas, i.e., \textit{energy harvesting and wireless energy transfer}, \textit{crowdsourcing}, and \textit{service selection and composition}. We describe the related work to our research in each of these domains. 
\vspace{-0.4 cm}
\subsection{IoT based crowdsourcing}
The mobile crowdsourcing~\cite{ren2015exploiting} has emerged recently as a new paradigm represented in a number of applications like crowdsencing and crowdcomputing. SignalGuru~\cite{koukoumidis2011signalguru} is a local-based mobile application to detect opportunistically current traffic signals by smartphones. These traffic signals are collobaratively exchanged through an ad-hoc network. The crowdcomputing also leverages mobile devices to compute outsourced data from different sources like sensors or smartphones. Honeybee~\cite{fernando2012honeybee} is a local-based application which outsources face detection tasks to the neighboring devices. Femtoclouds~\cite{ahabak2015femto} also represent significant exploitation of crowdsourcing smartphones capabilities. Femtoclouds are a cloud-type computing environment which harnesses a set of co-located smartphones as computing service providers for lightweight applications. CaaS~\cite{neiat2017crowdsourced} is a framework for WiFi hotspot sharing. This framework provides a WiFi covered trip plan by composing sensor-data services from public transportation and WiFi coverage services from public and individual hotspots.     
\vspace{-0.4cm}
 \subsection{Energy harvesting and wireless transfer}
In the context of IoT, body movement and heat provide a significant source of energy for wearables~\cite{choi2017wearable}. This energy can be converted to electric power which can satisfy IoT devices. Recently, several research have been conducted to integrate harvesting this energy into designing IoT objects~\cite{gorlatova2014movers,khalifa2017harke}. The Kinetic Energy Harvesting (KEH) for IoT is designed to capture kinetic energy from wearables by exercising different daily activities such as walking, running, and reading \cite{gorlatova2014movers}.
The advent of wireless charging makes the harvested energy from IoT devices more flexible and convenient to be easily shared. Energy sharing helps create self-sustained systems. Different techniques have been developed for the wireless charging in IoT and sensor networks \cite{lu2016wireless}. The most common techniques are magnetic inductive coupling, magnetic resonant coupling and microwave radiation. These techniques are used in wireless sensor networks by deploying charger robots in the network to charge the low battery sensors \cite{na2018energy}. A new paradigm of uncoupled wireless charging based on radio waves has emerged \cite{bell2019systems}. The Energous Wattup applies radio waves to enable wireless energy sharing for IoT devices. 

Wireless crowd charging is a new paradigm in the wireless transfer technology \cite{bulut2018crowdcharging,dhungana2019peer, raptis2019online}. This paradigm has been introduced by Bulut al. to provide IoT users with ubiquitous power access through crowdsourcing \cite{bulut2018crowdcharging}. Dhungana et al. provide a recent comprehensive survey on the use of peer-to-peer energy sharing in four different applications of mobile networks, namely, wireless sensor networks (WSN), mobile social networks (MSN), vehicular ad hoc networks (VANET) and UAV networks. They explore the technical approaches and mathematical tools adopted for the utilization of energy sharing in the aforementioned domains \cite{dhungana2019peer}. Raptis et al. claim that having even limited knowledge on the crowd network properties can be crucial for the design of crowd charging protocols. A key characteristic of such crowds is the active presence and involvement of the users in online social networks. They suggest the exploitation of online social information in order to tune the wireless crowd charging process \cite{raptis2019online}. In our paper, we harness the service paradigm to abstract wireless energy services and enable energy sharing in the crowdsourced IoT environment.
\vspace{-0.35cm}
\subsection{Service selection and composition}
The service selection and composition is a topical research in different domains such as cloud computing, sensor-cloud services, and social networks~\cite{neiat2017crowdsourced,zeng2004qos,ahabak2015femto}. In a sensor-cloud framework, services are composed according to their \textit{functional properties} and  the consumer preferences (QoS).The service composition methods usually convert the composition to a resource scheduling or an optimization problem. Resource scheduling in service composition has been extensively researched \cite{li2018service}. The fundamental parameters of resource scheduling algorithms are the optimization target and the scheduling priority.  The scheduling target in service composition is defined mostly by the resource utilization maximization, resource utilization fairness, or minimization of scheduling time \cite{yau2009adaptive, bae2007fairness}.  The scheduling priority defines which services to be privileged.  For example, the priority for Short Job First (SJF) scheduling algorithm is shortest jobs to be scheduled first \cite{ghanbari2012priority}. Similarly, Directed Acyclic Graphs (DAG) represent one of the major algorithms used in service composition to define the scheduling and priority policies of services. Optimization methods utilize different algorithms to obtain an optimal solution such as integer programming, genetic algorithm (GA), and particle swarm optimization (PSO) \cite{ai2011qos, ye2011genetic}. A data-driven service composition approach is proposed based on Petri-nets to meet the need of the business requirement~\cite{tan2010data}. Service functionalities may be constrained \cite{deng2016constraints}. Wang et al. proposes a constraint-aware service composition method \cite{wang2014constraint}. Their solution includes novel prepossessing techniques and a graph search-based algorithm~\cite{wang2014constraint}. Various QoS-aware service selection methods have been proposed in \cite{zeng2004qos}. Deng et al. proposed a service pruning method to address the QoS-correlation problem in service selection and composition ~\cite{deng2016service}.
The composition of crowdsourced services \cite{neiat2017crowdsourced} should consider two aspects, the spatio-temporal features of the consumers and their preferences. Usually, crowdsourced services are from different sources (mobile and static devices). The functional properties of the provided services should conform the spatio-temporal features of the query. The consumer preferences have to be met by the QoS of the provided services. Neiat et al.~\cite{neiat2017crowdsourced} designed a spatio-temporal service composition framework to compose WiFi hotspots services and to provide the most convenient trip plan with the best crowdsourced WiFi coverage. The proposed composition method uses the resulting composition of WiFi hotspots services as a QoS for selecting the best combination of line segment services.

The advancements in mobile devices and communication technologies enables the new mobile applications which rely on interaction between mobile services \cite{ahabak2015femto}. The traditional service composition techniques cannot perform in this pervasive mobile environment \cite{deng2016toward}.
Pervasive Information Communities Organization (PICO)~\cite{kumar2003pico} is one of the first proposed middleware to compose services in a pervasive environment. PICO  represents dynamic resources as services. The component services are modeled as directed attributed graphs. These basic services are dynamically combined to serve consumer requests. Han and Zhang developed a dynamic source routing based service composition protocol \cite{han2010design}. The proposed composition protocol efficiently consider QoS in real-time systems such as delay and cost. However, the developed composition protocol does not consider the users' mobility. A service composition technique that incorporates the providers' mobility is proposed in \cite{wang2010exploiting}. The proposed solution is efficient in term of selecting the most stable service in a dynamic environment. 

Crowdsourcing energy as a service is converted to a Qos-aware service composition problem. The spatio-temporal of energy services are considered as Qos attributes (i.e., start and end time, duration and location of an energy service). Composing energy services relies on finding the optimal selection of nearby services, which fulfils the required energy of a consumer within their query duration. Existing composition techniques may not be directly applicable to compose energy service due to the uniqueness of the crowdsourced IoT environment. An IoT user may consume only a part of the advertised energy from nearby services. Another key issue is the requirement of a novel composability model of energy services. The energy services should consider intensity compatibility between the user’s IoT device and the providing devices in a composition.

\vspace{-0.4cm}
\section{Conclusion}
We propose a novel composition framework to crowdsource wireless energy services from IoT devices. We design a novel composability model considering the energy usage behavior and the spatio-temporal aspects of the IoT devices. We formulate the composition problem as a multi-objective optimization of meeting users' energy requirements in the earliest and shortest time intervals. We conduct a set of experiments to investigate and compare the scalability and the effectiveness of the proposed composition techniques, i.e., a greedy composition technique, the multi-knapsack composition and a heuristic based composition approach. In an IoT environment, the energy services might scale from very small capacity provided by tiny devices to considerable capacity provided by bigger devices. Results show that the proposed approach is scalable and effective in various composition scenarios. Experiment results depict that the greedy and heuristic based composition approaches are more scalable and runtime efficient than the the multi-knapsack composition approach. In the future work, we will explore the mobility challenges for composing services in a crowdsourced IoT environment.
\section*{Acknowledgement}
This work was also made possible by NPRP9-224-1-049 grant from the Qatar National Research Fund (a member of Qatar Foundation) and DP160100149 and LE180100158 grants from the Australian Research Council. The statements made herein are solely the responsibility of the authors.
\vspace{-0.5cm}
\bibliographystyle{IEEEtran}
\bibliography{IEEEabrv,bibShort}
\vspace{-1.3cm}
\begin{IEEEbiography}[{\includegraphics[width=1in,height=1.25in,clip,keepaspectratio]{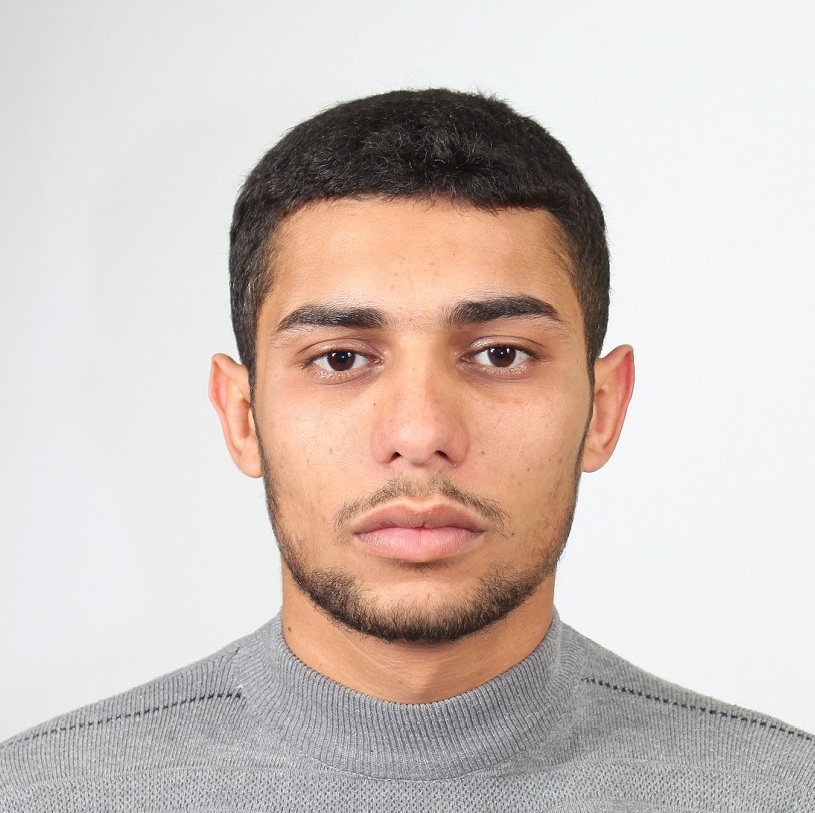}}]{Abdallah Lakhdari} 
is a PhD candidate under the supervision of professor Athman Bouguettaya. He spent one year in Cloud Computing and Big Data “C2BD” lab at New Mexico tech, USA as a visiting graduate student. Abdallah received his bachelor degree and Masters degree in Computer Science from University of Laghouat, Algeria in 2010 and 2013 respectively. 
His research interests include Social computing, crowdsourcing, IoT, and Service Computing.
\end{IEEEbiography}
\vspace{-1.3 cm}
\begin{IEEEbiography}[{\includegraphics[width=1in,height=1.25in,clip,keepaspectratio]{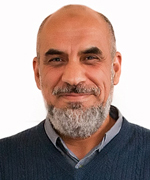}}]{Athman Bouguettaya}
is Professor and Head of School of Computer Science at University of Sydney, Australia. He received his PhD in Computer Science from the University of Colorado at Boulder (USA) in 1992. He is or has been on the editorial boards of several journals including, the IEEE Transactions on Services Computing, ACM Transactions on Internet Technology, the International Journal on Next Generation Computing and VLDB Journal. He is a Fellow of the IEEE and a Distinguished Scientist of the ACM.
\end{IEEEbiography}
\vspace{-1.3cm}
\begin{IEEEbiography}[{\includegraphics[width=1in,height=1.25in,clip,keepaspectratio]{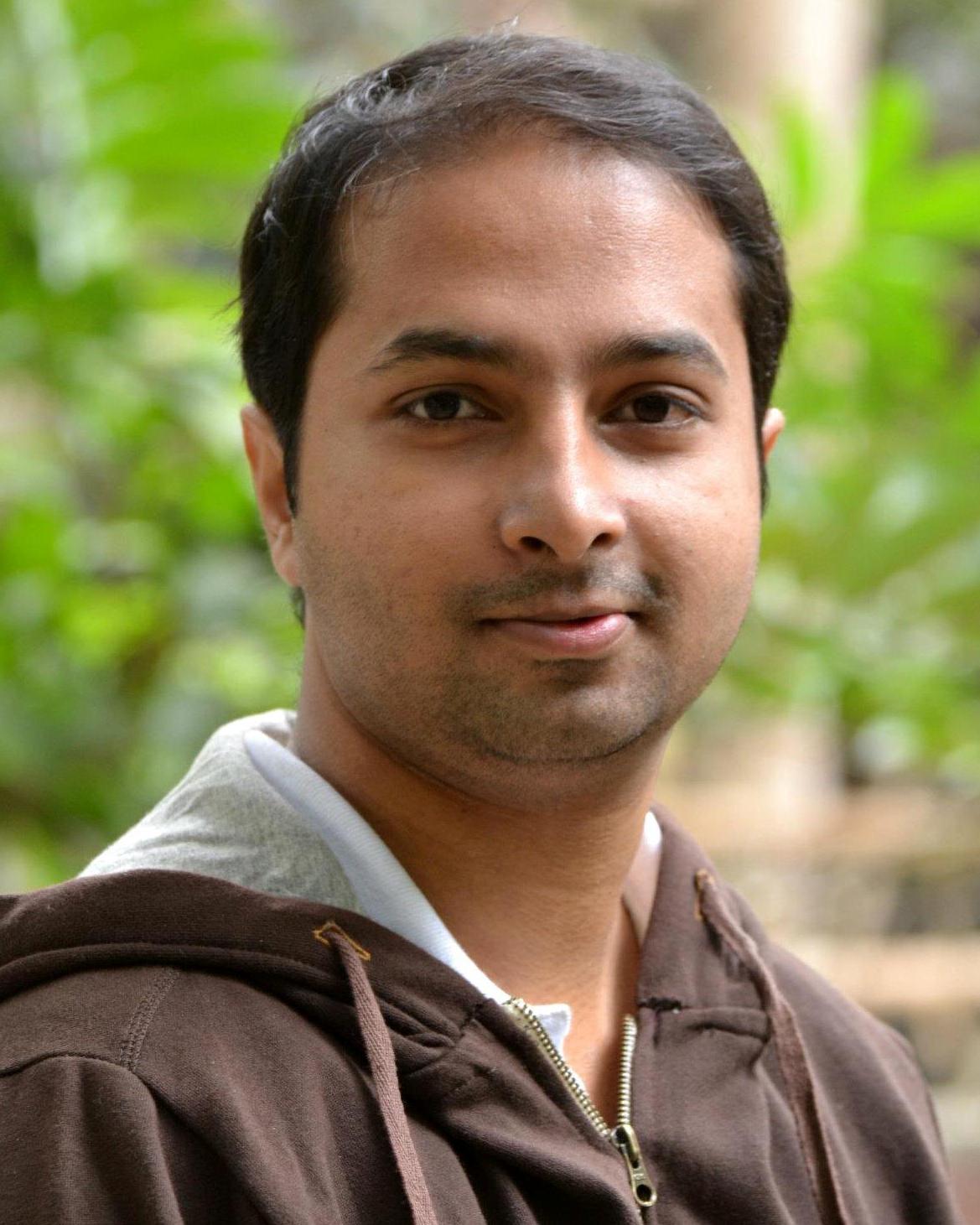}}]{Sajib Mistry} 
is Lecturer at School of Elect Eng, Computer and Math Sci in Curtin University, Australia. He was a postdoc fellow at School of Computer Science in University of Sydney. He received his PhD from the RMIT University, Australia in 2017. His research interests include Edge/Cloud Computing, Big Data and IoT. He has published articles in international journals and conferences, such as IEEE TSC, TKDE, ACM CACM, ICSOC, WISE, and ICWS. He received the best paper award in ICSOC2016. 
\end{IEEEbiography}
\vspace{-1.3cm}
\begin{IEEEbiography}[{\includegraphics[width=1in,height=1.25in,clip,keepaspectratio]{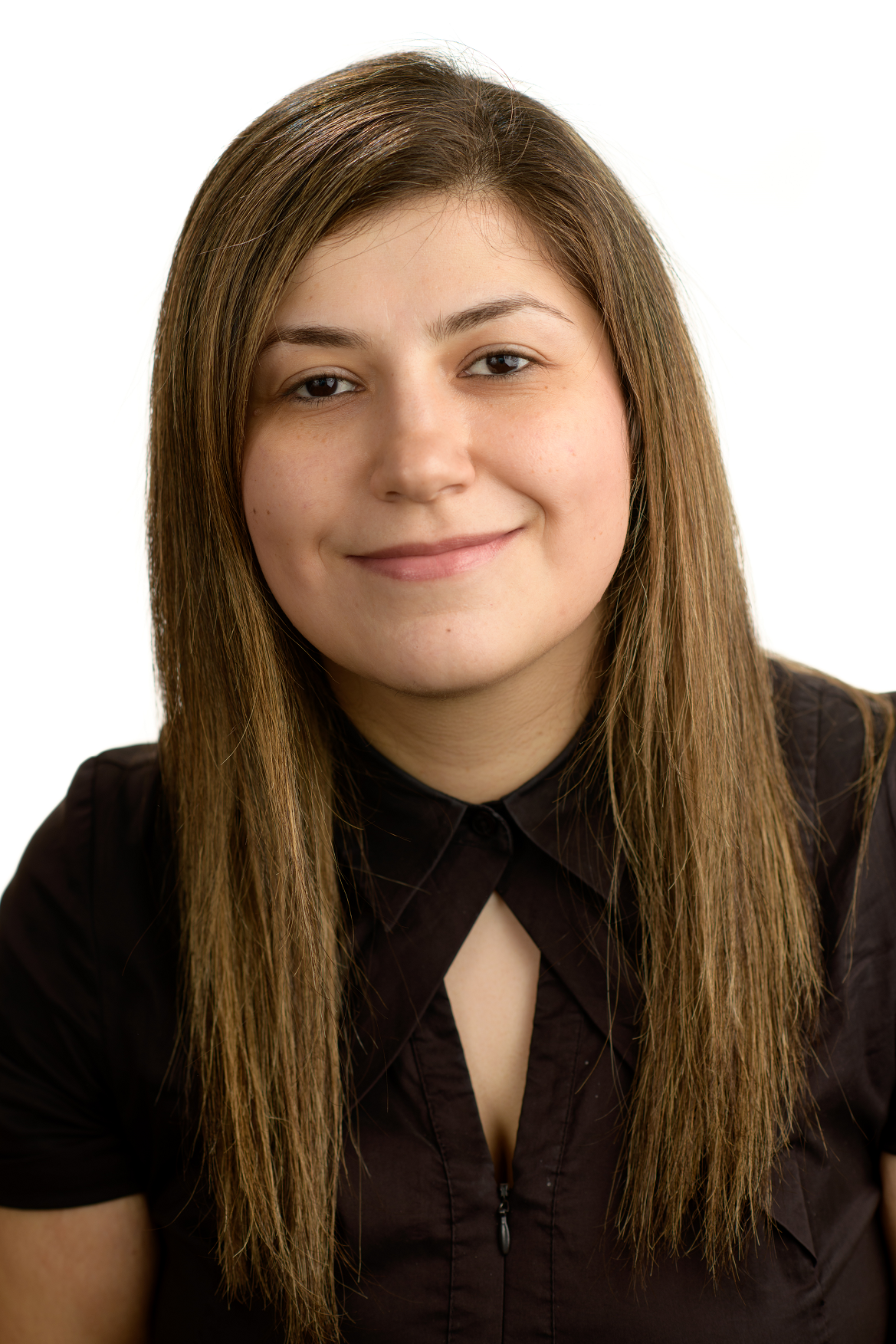}}]{Azadeh Ghari Neiat}
is a lecturer in the School of Information Technology at the Deakin University. She was awarded a PhD in computer science at RMIT University, Australia in 2017. Her current research interests include Internet of Things (IoT), Spatio-Temporal Data Analysis, Mobile Crowdsourcing/Crowdsensing, Big Data Mining, and Machine Learning with applications in the smart city, smart home, and recommender systems. 
\end{IEEEbiography}

\end{document}